\documentstyle[12pt,epsf]{article}      
\newcommand{\OO}{{\cal O}}
\newcommand{\HH}{{\cal H}}

\newcommand{\TT}{{\cal T}}

\newcommand{\wt}{\widetilde}
\newcommand{\wh}{\widehat}

\newcommand{\be}{\begin{equation}}
\newcommand{\ee}{\end{equation}}
\newcommand{\ben}{\begin{eqnarray}\displaystyle}
\newcommand{\een}{\end{eqnarray}}
\newcommand{\refb}[1]{(\ref{#1})}
\newcommand{\p}{\partial}
\newcommand{\sectiono}[1]{\section{#1}\setcounter{equation}{0}}

\newcommand{\hp}{{\wh\Phi}}
\newcommand{\hq}{{\wh Q_B}}
\newcommand{\he}{{\wh\eta_0}}
\newcommand{\ha}{{\wh{A}}}
\newcommand{\lll}{\big\langle\big\langle}
\newcommand{\rrr}{\big\rangle\big\rangle}
\newcommand{\lllb}{\Bigl\langle\Bigl\langle}
\newcommand{\rrrb}{\Bigr\rangle\Bigr\rangle}

\begin{document}

{}~
\hfill\vbox{\hbox{hep-th/0002211}\hbox{CTP-MIT-2951}
\hbox{IFT-P.018/2000} \hbox{MRI-PHY/P20000203}
}\break

\vskip 1.8cm

\centerline{\large \bf Tachyon Condensation in Superstring Field Theory}

\vspace*{6.0ex}

\centerline{\large \rm Nathan Berkovits$^a$, Ashoke Sen$^b$ and Barton 
Zwiebach$^c$}

\vspace*{6.5ex}

\centerline{\large \it ~$^a$Instituto de F\'{\i}sica Te\'orica, Universidade
Estadual Paulista}

\centerline{\large \it Rua Pamplona 145, 01405-900, S\~ao Paulo, SP, BRASIL}

\centerline{E-mail: nberkovi@ift.unesp.br}

\vspace*{1ex}

\centerline{\large \it ~$^b$Mehta Research Institute of Mathematics
and Mathematical Physics}

\centerline{\large \it  Chhatnag Road, Jhoosi,
Allahabad 211019, INDIA}

\centerline{E-mail: asen@thwgs.cern.ch, sen@mri.ernet.in}

\vspace*{1ex}

\centerline{\large \it $^c$Center for Theoretical Physics, 
Massachussetts Institute of Technology}

\centerline{\large \it Cambridge,
MA 02139, USA}

\centerline{E-mail: zwiebach@mitlns.mit.edu}

\vspace*{4.5ex}

\centerline {\bf Abstract}
\bigskip
It has been conjectured that at the stationary point of the tachyon
potential for the D-brane-anti-D-brane pair or for the non-BPS D-brane
of superstring theories, the negative energy density cancels the
brane tensions. We study this conjecture using a Wess-Zumino-Witten-like
open superstring field theory free of contact term divergences and 
recently shown to give 60\% of the vacuum energy by condensation of 
the tachyon field alone.   While the action is non-polynomial,
the multiscalar tachyon potential to any  fixed level involves only 
a finite number of interactions.   We compute this potential to 
level three, obtaining 85\% of the expected vacuum energy, a result 
consistent with convergence that can also be viewed as a successful 
test of the string field theory. The resulting effective tachyon 
potential is bounded below and  has two degenerate global minima. 
We calculate the energy density of the kink solution interpolating 
between these minima finding good agreement with the tension of 
the D-brane of one lower dimension.

\vfill \eject

\baselineskip=18pt

\tableofcontents

\sectiono{Introduction} \label{s0}

The spectrum of open strings living on a  
D-brane anti-D-brane pair of type IIA or IIB string theory contains a pair
of tachyonic modes from the Neveu-Schwarz (NS) sector, 
indicating that the system is
unstable \cite{9511194,9604091}.  
There are
general arguments \cite{9805019,9805170,9808141} which indicate that the
tachyonic potential has a
minimum, and this minimum represents the usual vacuum of the closed string
theory without any D-brane.\footnote{Some of these arguments used
earlier gauge theory analysis of 
brane-antibrane systems, see \cite{9704006,9703217}.}
For this to be true, the negative  
energy density 
contribution from the tachyon potential 
at the minimum  must
exactly cancel the sum of the tensions of the brane-antibrane pair.

During the last few years
it has been realised that type IIA (IIB) string theory contains unstable 
non-BPS D$p$-branes for odd (even) 
$p$ \cite{9806155,9808141,9809111,9812031,9812135,0001143}. Any of these 
D-branes
has a tachyonic mode, indicating that the brane is unstable. A
straightforward extension of the general arguments for the brane-antibrane
system can be used to argue that the tachyonic potential has a minimum,
and this minimum represents the usual vacuum of the closed string theory
without any D-brane. For this to be true, the negative 
energy density 
contribution from the tachyon potential  
at this minimum must exactly
cancel the tension of the non-BPS D-brane.

A similar conjecture exists also for D-branes of bosonic string
theory \cite{RECK,9902105}. On any
bosonic string D-brane   
there is a tachyonic open string 
mode. Indirect arguments, similar to those for the
brane-antibrane pair of type II string theories, indicate that the tachyon
potential
has an extremum whose negative 
energy density contribution 
cancels the tension of the D-brane, so that this particular extremum
represents the vacuum of closed bosonic string theory without any
D-branes. In refs.\cite{9911116,9912249} this phenomenon was studied
directly in open bosonic string
field theory \cite{WITTENBSFT}, following earlier work of ref.\cite{KS}.
Using the level truncation scheme of ref.\cite{KS}, 
ref.\cite{9912249} showed that including scalars up to
level four, the value of the potential at the extremum
cancels
almost 99\% of the D-brane tension. This is a strong
indication that this extremum indeed represents vacuum without
D-branes. This remarkable
cancellation has now been verified to an accuracy of 
99.9\% by including scalars up to level ten \cite{moeller}.  
Evidence to the validity of the level expansion
was recently given in \cite{0001201}. Further evidence 
for the identification of the
tachyonic vacuum has been found in ref.\cite{0002117} 
who considered tachyonic lump solutions of the string equations of motion.

In a recent paper \cite{0001084}, the zeroth order contribution to the
tachyon potential on a non-BPS D-brane of type II string theory was
computed 
using the open string field theory action formulated in
refs.\cite{9503099,9912121},
and was found to contain
a minimum at which the potential cancels 60\% of the D-brane tension.
Unlike the cubic 
action proposed in \cite{WITTENSFT},
the Wess-Zumino-Witten-like
action used in ref.\cite{0001084} has no problems with contact term
divergences \cite{9912120}.\footnote{An early  
attempt at generalizing 
the analysis of ref.\cite{KS} to superstring theory was made in
ref.\cite{AREF}.} 
Although it is not known how to include Ramond (R) 
sector states
into this action in a manifestly SO(9,1) covariant manner,
this is not a problem  here 
since the phenomenon of tachyon condensation involves 
NS sector states only. 
Of course, it involves the full unprojected NS sector, namely
both GSO$(+)$ and GSO$(-)$ states. 
While there is a superspace version of the Wess-Zumino-Witten-like
action which is manifestly SO(3,1) super-Poincar\'e invariant
and includes all GSO$(+)$ states in the NS and R sectors,
it is not yet known how to incorporate GSO$(-)$ states
into it.

In this paper, we compute the first-order correction
to the tachyon potential on a non-BPS D-brane of type II string theory 
using the same action as \cite{0001084}, 
and find a minimum of the potential at which
85\% of the D-brane tension is cancelled by the potential. This provides
strong evidence that
the approximation scheme is converging as in the bosonic
string computation and that 
the exact tachyon potential has a minimum where the D-brane tension
is exactly cancelled. Alternatively, this result can
also be viewed as a successful
test of the correctness of this
superstring field theory action.

Although we
carry out the explicit analysis for the non-BPS D-brane, the result
also holds for the brane-antibrane system. Indeed, the tachyon potential
on the non-BPS D-brane of type IIA (IIB) string theory can be obtained
from the tachyon potential on a brane-antibrane system of type IIB (IIA)
string theory after restricting the field configuration to a $Z_2$
invariant subspace\cite{9812031,0001143}; so the existence of a minimum of
the tachyon potential for a non-BPS D-brane corresponding to the vacuum
without D-branes also establishes the corresponding result for the
brane-antibrane system. This can also be made self-evident by comparing
the
structure of the string field theory action on the non-BPS D-brane and
that on a brane-antibrane pair, both of which we write down explicitly.

Since the tachyon potential on a non-BPS D-brane is invariant under a
change of sign of the tachyon field, there are doubly degenerate minima of
the potential, and we can construct a kink solution interpolating between
these two minima. It has been conjectured that this represents a BPS brane
of one lower dimension\cite{9812031,9812135} following a similar
conjecture for the brane-antibrane system\cite{9805019,9808141}. We
compute numerically  
the energy density of the kink solution using the
tachyon potential, but ignoring string field theory corrections to the
tachyon kinetic term, in the same spirit as in a recent paper for bosonic
string field theory \cite{0002117}. The result is 1.03 times the expected
answer. Although such a close agreement is likely to be accidental, it is
encouraging to note that the mass of the kink even in this crude
approximation has the correct order of magnitude. We should also note that
the effect of the non-zero tachyon background should be to 
reduce the kinetic
term, since at the minimum of the potential the kinetic term is expected
to vanish, so that we have no  physical 
excitations. Thus we expect that
once we take into account corrections 
to the kinetic term, the energy of
the kink should be lowered. In this context, it is also encouraging
to note that in the analysis of ref.\cite{0002117} the mass of the lump
{\it decreased} after taking into account corrections 
to the tachyon kinetic energy term. 

\medskip   
The paper is organised as follows.
In section \ref{s1} of this
paper, we shall review this WZW-like action, discuss in detail its
basic ingredients, 
its gauge invariance, and its application to describe the non-BPS D-brane
as well as the brane anti-brane system.  In section
\ref{s2}, we
shall use the action to compute the zeroth and first order contributions
to the tachyon potential and show that the potential has
a minimum at 85\% of the D-brane tension.  In section \ref{s4} we discuss
the
tachyonic kink solution and calculate its mass. We offer some 
perspective on our results and discuss open questions in section \ref{s5}. 
Important details have been provided in the appendices.  Appendix A
establishes the cyclicity properties of the amplitudes appearing in
the string action-- this cyclicity is  essential for gauge invariance.
Appendix B explains the twist properties of the amplitudes-- such properties
allows us to restrict the multiscalar tachyon field (the space ${\cal
H}_1$ defined in section \ref{s2})
to the twist even
sector.  Appendix C gives a self-contained derivation of the mass
of the D-brane described by a string field theory action.  Finally,
in appendix D we provide details on the computation of the 
tachyon potential.

\sectiono{Open superstring field theory} \label{s1}

In this section we shall explain and analyze the superstring 
field theory that describes  the dynamics
of a non-BPS D-brane of type II string theory. 
As it will be clear, this string field theory
is readily 
modified  to discuss the D-brane anti-D-brane system in superstring
theory. In fact, the same calculations give the tachyon potential
for both physical situations.  As in the case of 
refs.\cite{9911116,9912249,0001084}, we shall not restrict our analysis 
to any specific background,
but will assume, for convenience, that all the directions 
tangential to the D-brane are
compact, so that the system has a finite mass. 

We will begin by discussing the GSO projected, or GSO$(+)$ sector
of the  open superstring theory formulated in refs.\cite{9503099,9912121},
$-$ this would describe the dynamics of
NS sector open strings living on a single BPS D-brane. Here the {\it basic
structure} of the theory will be elaborated. Then we
turn to the non-BPS D-brane whose formulation requires
incorporating both the GSO$(-)$ sector and the GSO$(+)$ sector
of the theory. This can be done by attaching {\it internal}
Chan-Paton matrices to the GSO plus and minus sectors
in such a way that the complete string field and the
relevant operators satisfy the basic structure of the
original GSO$(+)$ theory. This device was used
in \cite{0001084}  for the analysis of the non-BPS D-brane.
Finally, in the last subsection
we show how, in addition to the internal Chan-Paton matrices,
{\it external} Chan-Paton matrices must be tensored to 
describe the brane-antibrane system.   

\subsection{Superstring field theory on a BPS D-brane} 

In the formalism of refs.\cite{9503099,9912121,0001084}, a general
off-shell string field
configuration in the GSO$(+)$ NS sector corresponds to 
a {\it Grassmann even} open string vertex operator
$\Phi$ of ghost number 0
and picture number\cite{FMS} 0 in the combined
conformal field theory of a $c=15$ superconformal matter system, and
the $b,c,\beta,\gamma$ ghost system with $c=-15$. In terms of the
bosonized ghost fields $\xi,\eta,\phi$ related to $\beta,\gamma$ through
the relations 
\be \label{ei1}
\beta=\p\xi e^{-\phi}, \qquad \gamma =\eta e^\phi,
\ee
the ghost number ($n_g$) and the picture number ($n_p$) assignments are as
follows:
\ben \label{ei2}
&& b:\quad n_g=-1, n_p=0, \qquad c:\quad n_g=1, n_p=0 \, , \nonumber \\
&& e^{q\phi}:\quad n_g=0, n_p=q\, , \nonumber \\
&& \xi:\quad n_g=-1, n_p=1, \qquad \eta:\quad n_g=1, n_p=-1\, . \nonumber
\\
\een
The SL(2,R) invariant vacuum carries zero ghost and picture number.
Note that this definition of ghost number agrees with the
definition of \cite{FMS} for states with zero picture, but unlike
the definition of \cite{FMS}, it allows the spacetime-supersymmetry
generators to carry zero ghost number\cite{9912121}.
One notable difference
from other formulations of open string field theory is that here the
string field correspond to vertex operators in the `large Hilbert space'
containing the zero mode of the field $\xi$. 

We shall denote by $\langle \prod_i A_i \rangle$ the correlation function
of a set of vertex operators in the combined matter-ghost conformal field
theory on the unit disk
with open string vertex operators inserted on the boundary of the disk,
without including trace over CP factors.
These
correlation functions are to be computed with the
normalization
\be \label{eb1}
\langle\xi(z) c\p c\p^2 c(w) e^{-2\phi(y)}\rangle = 2\, .
\ee
Throughout this paper we shall be working in units where  $\alpha'=1$.
The nilpotent BRST operator of this theory is given by
\be \label{eb4}
Q_B = \oint dz j_B(z)
= \oint dz \Bigl\{  c \bigl( T_m + T_{\xi\eta} + T_\phi)
+ c \partial c b +\eta \,e^\phi
\, G_m - \eta\p \eta e^{2\phi} b \Bigr\}\, ,
\ee
where
\be \label{eb5}
T_{\xi\eta}=\p\xi\,\eta, \quad T_\phi=-{1\over 2} \p\phi \p
\phi -\p^2\phi \, ,
\ee
$T_m$ is the matter stress tensor and 
$G_m$ is the matter
superconformal generator. $G_m$ is a dimension $3/2$ primary field and 
satisfies:
\be \label{eb6}
G_m(z) G_m(w) \simeq  {10\over (z-w)^3} + {2T_m\over (z-w)}\, .
\ee
The
normalization of
$\phi$, $\xi$, $\eta$, $b$ and $c$ are as follows:
\be \label{eb3}
\xi(z) \eta(w) \simeq {1 \over z-w}, \quad b(z) c(w)\simeq {1\over z-w}, 
\quad \p \phi(z)
\p\phi(w)\simeq
-{1\over (z-w)^2} \, .
\ee
We denote by $\eta_0=\oint dz \eta(z)$ the zero mode of 
the field $\eta$ acting on the Hilbert space of matter ghost CFT.

The string field theory action  is
given by\cite{0001084}
\be \label{e0}
S={1\over 2g^2}\lllb (e^{-\Phi} Q_B e^{\Phi}) 
(e^{-\Phi}\eta_0 e^\Phi)
- \int_0^1 dt 
(e^{-t\Phi}\p_t e^{t\Phi})\{ (e^{-t\Phi}Q_B e^{t\Phi}),
(e^{-t\Phi}\eta_0 e^{t\Phi})\}\rrrb\, , 
\ee 
where $\{ A, B\} \equiv AB+ BA$, and 
$e^{-t\Phi}\p_t e^{t\Phi} = \Phi$ but has 
been written this way for convenience. This action is defined  
by expanding all exponentials in formal Taylor series carefully
preserving the order of all operators and letting $\lll\cdots \rrr$
of an ordered sequence of arbitrary vertex operators $A_1, \ldots A_n$
be defined as:
\be \label{e2ff}
\lll A_1\ldots A_n \rrr = \Bigl\langle f^{(n)}_1 \circ A_1(0)\cdots 
f^{(n)}_n\circ A_n(0)\Bigr\rangle\, .
\ee
Here, 
$f\circ A$ for
any function $f(z)$, denotes the conformal transform of $A$ by $f$, and 
\be \label{e3}
f^{(n)}_k(z) = e^{2\pi i (k-1)\over n} \Big({1+iz\over 1-iz}
\Big)^{2/n}\,\quad  \hbox{for} \quad n\geq 1 .   
\ee
 In
particular if
$\varphi$
denotes a primary field of weight
$h$, then
\be \label{e3b}
f\circ \varphi(0) = (f'(0))^h \varphi(f(0))\, .
\ee 
$Q_B$ (or $\eta_0$) acting on a set 
of vertex operators inside $\lll~\rrr$
corresponds to a contour integral of $j_B$ (or $\eta$) around the
insertion points of these vertex operators on the right hand side of
eq.\refb{e2ff}.

Since we have, in general, non-integer weight vertex operators, we
should be more careful in defining $f\circ A$ for such vertex
operators.
Noting that 
\be \label{e3c}
f^{(N)\prime}_k(0) = {4 i\over N} e^{2 \pi i{k-1 \over N}} \equiv {4\over
N}
e^{2\pi i ({k-1\over N} + {1\over 4})}\, ,
\ee
we adopt the following definition of $f^{(N)}_k\circ\varphi(0)$ for a
primary
vertex operator $\varphi(x)$ of conformal weight $h$:
\be \label{e3d}
f^{(N)}_k\circ\varphi(0) = \bigg|\Big({4\over N}\Big)^h\bigg| e^{2\pi i h
({k-1 \over N}+{1\over
4})}\varphi(f^{(N)}_k(0))\, .
\ee
Since all secondary vertex operators can be obtained as products of
derivatives of primary vertex operators, this uniquely defines
$f^{(N)}_k\circ A(0)$ for all vertex operators.

The geometry of the
interaction described in \refb{e3} is simple. 
The function $f^{(n)}_1$ maps  the  
upper half disk $|z|\leq 1, \Im(z) >0$ into the 
wedge $|\hbox{Arg}(f^{(n)}_1)|\leq \pi/n,\,|f^{(n)}_1| \leq 1$, with
the puncture $z=0$ ending at $f^{(n)}_1=1$. With $k=1, \cdots n$, we 
end up gluing $n$ such wedges together to form a full unit disk where
the $n$ vertex operators are inserted at equally spaced points on
the boundary.
By a further
SL(2,C) transformation $F$ ({\it e.g.} $F(w)=i(1-w)/(1+w)$) we can map
the interior of the unit disk onto the upper half plane. We could use the
functions $g^{(n)}_k(z)=F(f^{(n)}_k(z))$ instead of $f^{(n)}_k(z)$ to
define the string field theory action. $\langle~\rangle$ will now denote
the correlation function of the conformal field theory on the upper half
plane, with open string vertex operators inserted on the  real
axis. As will be shown in appendix \ref{a3}, by a convenient choice of the
SL(2,C) transformation $F$ we can
ensure that $g^{(n)}_1(0), \ldots g^{(n)}_n(0)$ are ordered from left to
right on the real axis. Also one finds that $g^{(n)\prime}_k(0)$ is real
and positive for all $k$. The prescription \refb{e3d} then corresponds to
choosing real, positive values of $(g^{(n)\prime}_k(0))^h$ in the
expression for the conformal transform of a field $\Phi$ of weight $h$. 
As a double-check of our computations, we shall compute the tachyon
potential using both the disk and the UHP prescriptions and compare
answers. 
 
\medskip
The correlator $\lll ~ \rrr$ defined in eq.\refb{e2ff} 
satisfies cyclicity properties.
Let $ \Phi$ denote any component of the string field,
and $A_1,\ldots
A_{n-1}$
denote
arbitrary  vertex operators ({\it i.e.} arbitrary grassmanality,
ghost number, etc.). Then
\ben \label{e1aa}
&& \lll A_1\ldots A_{n-1} \Phi \rrr = 
\lll \Phi A_1\ldots A_{n-1}
\rrr, \nonumber \\
&& \lll A_1\ldots A_{n-1} (Q_B\Phi) \rrr = -
\lll (Q_B\Phi) A_1\ldots A_{n-1} \rrr, \nonumber \\
&& \lll A_1\ldots A_{n-1}
(\eta_0\Phi) \rrr = -
\lll (\eta_0\Phi) A_1\ldots A_{n-1} \rrr\, .
\een
The proof of these relations has been given in appendix \ref{a1}.

Note that in this notation the BPZ inner product is given by:
\be \label{ebpz}
\langle A|B\rangle = \lll A B \rrr\, ,
\ee
which uses the two punctured disk (eqn.\refb{e3} with $n=2$). 
We now define the multilinear products 
$|A_1A_2\ldots A_n\rangle$ of $n$ vertex operators
$A_1, A_2,\ldots A_n$ through the relation:
\be \label{estar}
\langle B| A_1\ldots A_n\rangle = \lll B A_1\ldots A_n\rrr\, , 
\ee
for any state $\langle B|$. The product $|A_1A_2\rangle$, computed
with  \refb{e3} and  $n=3$, is simply the associative 
(non-commutative) star product
$|A_1* A_2\rangle$ of \cite{WITTENBSFT}.  It follows from the
geometry of the interaction that the higher products are 
equivalent to iterated multiplication using the star product:
namely, $|A_1A_2\cdots A_n\rangle = |A_1*A_2*\cdots *A_n\rangle$.
While the order of the sequence of operators must be preserved, 
the  multiplications in this ket can be done in
any order, thanks to the associativity of the star product. It  follows
that all products associate. 
{}From now on we shall denote the 
product of a set of vertex operators $A_1,
A_2,\ldots A_n$ by $A_1A_2\ldots A_n$.

It will now be shown that 
this action is invariant under the gauge transformation\cite{0001084}
\be \label{egtrs}
\delta e^{\Phi} = (Q_B \Omega) e^{\Phi} + e^{\Phi}(\eta_0\Omega') \, , 
\ee
where the gauge transformation parameters $\Omega$ and $\Omega'$ are
Grassmann   odd, 
GSO$(+)$ vertex operators with $(n_g,n_p)$ values $(-1,0)$ and $(-1,1)$
respectively. 
The proof will use the cyclicity relations \refb{e1aa} and the following
identities:
\be \label{epr1}
\{ Q_B, \eta_0\} = 0, \quad Q_B^2 = \eta_0^2 = 0 ,
\ee
\be \label{epr2}
Q_B (\Phi_1 \Phi_2) = (Q_B\Phi_1)\Phi_2 + 
\Phi_1 (Q_B\Phi_2),\quad
\eta_0 (\Phi_1\Phi_2) = 
(\eta_0\Phi_1)\Phi_2 + \Phi_1 (\eta_0\Phi_2), 
\ee
\be \label{epr3}
\lll Q_B (...) \rrr =
\lll \eta_0(...) \rrr =0.
\ee
Note that in the identities of the
second line there are no minus signs necessary as $Q_B$ or 
$\eta_0$ go through the string field because the string
field is Grassmann even. The identities in the  last
last line hold because $Q_B$ and $\eta_0$
are integrals of
dimension one currents which can be ``pulled'' off the 
boundary and collapsed inside the disk. 

Defining $G= e^{\Phi}$ and using the above identities,
one finds that under an arbitrary variation
$\delta G$, 
\be \label{varia}
\delta S = {1\over { g^2}}\lll G^{-1}\delta G  \eta_0 (G^{-1} Q_B G) 
\rrr
\ee
where the first term of $S$ contributes 
$(2g^2)^{-1}G^{-1}\delta G [ \eta_0(G^{-1}Q_B G)-   
Q_B (G^{-1}\eta_0 G) ] $ to the variation
and the second term of $S$ contributes
$(2g^2)^{-1} G^{-1}\delta G [ \eta_0(G^{-1}Q_B G)
+Q_B(G^{-1}\eta_0 G) ] $  
to the variation.
Using the fact that $S$ goes to $-S$ after switching $\eta_0$
with $Q_B$ and $G$ with $G^{-1}$, \refb{varia} can also be written as 
\be \label{variatwo}
\delta S = -{1\over { g^2}}\lll G\delta G^{-1}  Q_B(G\eta_0 G^{-1})  
\rrr
\ee
To prove gauge invariance under
$\delta G= G\eta_0\Omega'$, use \refb{varia} and pull $\eta_0$ off
the $(G^{-1}Q_B G)$ term. Since $\eta_0(G^{-1}\delta G)=0$, $\delta S=0$.
To prove gauge invariance under 
$\delta G = (Q_B \Omega) G$, use \refb{variatwo} and pull
$Q_B$ off of the $(GQ_B G^{-1})$ term. Since 
$Q_B(G\delta G^{-1})=-Q_B (\delta G G^{-1})=0$, $\delta S=0$.
So we have proven invariance of the action 
under the gauge transformations of
\refb{egtrs}. 

The equation of motion for the action is easily
derived from \refb{varia} to be
\be\label{eom}
\eta_0(e^{-\Phi}Q_B e^{\Phi}) =0.
\ee

As stated earlier, the string field in the present theory
corresponds to vertex operators in the `large Hilbert space' which includes
the zero mode of $\xi$. However, using the $\Omega'$ gauge invariance, we
can choose the gauge $\xi_0\Phi=0$. In that gauge, the string field
configuration $\Phi$ is in one to one correpondence with
vertex operators in the `small Hilbert space' which does not include the
zero mode of $\xi$. This will be discussed in some detail in section
\ref{s2}.

\subsection{Superstring field theory on a Non-BPS D-brane} 

The open string states living on a single
non-BPS D-brane are  divided into two classes, GSO$(+)$ states
and GSO$(-)$ states.  Since the GSO$(-)$ states are
{Grassmann odd} they cannot be incorporated directly
into a string field preserving the algebraic structure
reviewed in the previous subsection. This structure
can be recovered by tensoring $2\times 2$ matrices 
carrying
{\it internal} Chan-Paton (CP) indices.
These are added both to the vertex operators and
to $Q_B$ and $\eta_0$.

We attach the  $2\times 2$ identity matrix $I$ 
on the usual GSO$(+)$ sector (recall that the
Neveu-Schwarz (NS) sector ground state is odd under the projection
operator $(-1)^F$) and  the
Pauli matrix $\sigma_1$ to the  GSO$(-)$ sector. The complete
string field is thus written as
\be
\label{fsf}
\hp = \Phi_+ \otimes I  + \Phi_- \otimes \sigma_1\,,
\ee
where the subscripts denote the $(-)^F$ eigenvalue of the
vertex operator.   In addition, we define:
\be \label{ei3}
\wh Q_B = Q_B\otimes \sigma_3, \qquad
\wh \eta_0 = \eta_0\otimes \sigma_3\, .
\ee
Note that this definition shows that these matrices do
not really carry conventional CP indices;  had it been 
so, both
$Q_B$ and $\eta_0$ should have been tensored with the
identity matrix, as such operators should not change the sector
the strings live in \cite{9705038}.  Finally, we define
\be \label{e200}
\lll \wh A_1\ldots \wh A_n \rrr = Tr \Bigl\langle f^{(n)}_1 \circ
\ha_1(0)\cdots
f^{(n)}_n\circ \ha_n(0)\Bigr\rangle \, ,
\ee
where the trace is over the internal CP matrices.
We shall adopt the convention that fields or operators with internal
CP factors
included are denoted by symbols with a
hat on them, and fields or operators without internal
CP factors included are
denoted by symbols without a hat, as in the previous subsection. 

Indeed, with these definitions the cyclicity relations \refb{e1aa}
given in the previous section now hold as
\ben \label{e1aaa}
&& \lll \ha_1\ldots \ha_{n-1} \hp \rrr = \lll \hp \ha_1\ldots \ha_{n-1}
\rrr, \nonumber \\
&& \lll \ha_1\ldots \ha_{n-1} (\hq\hp) \rrr = -
\lll (\hq\hp) \ha_1\ldots \ha_{n-1} \rrr, \nonumber \\
&& \lll \ha_1\ldots \ha_{n-1}
(\he\hp) \rrr = -
\lll (\he\hp) \ha_1\ldots \ha_{n-1} \rrr\, ,
\een
where $\wh \Phi$ denotes any component of the the string field,
and $\ha_1,\ldots
\ha_{n-1}$
denote arbitrary vertex operators. 
The proof of these relations, as well as those for
the unhatted case have been given in appendix \ref{a1}.
In addition, we have the analogs of \refb{epr1} holding
\be \label{epr11}
\{\hq ,\he\} = 0, \quad  \hq^2 = \he^2 =0,
\ee
\be \label{epr2a} 
\hq (\wh\Phi_1 \wh\Phi_2) = (\hq\wh\Phi_1)\wh\Phi_2 + 
\wh\Phi_1 (\hq\wh\Phi_2),\quad
\he (\wh\Phi_1 \wh\Phi_2) = 
(\he\wh\Phi_1)\wh\Phi_2 + \wh\Phi_1 (\he\wh\Phi_2), 
\ee
\be \label{epr3a}
\lll \hq(...) \rrr =
\lll \he(...) \rrr =0.
\ee
The reason no extra signs appear in the middle equation is clear,
when the string field is Grassmann odd  
the sign arising by moving $Q_B$ across
the vertex operator is cancelled by 
having to move $\sigma_3$ across $\sigma_1$.

Given that the relations satified by the hatted objects
are the same as those of the unhatted ones, 
the string field  action for the non-BPS D-brane takes the  
same structural form as that in \refb{e0} and is given by\cite{0001084}
\be \label{e00}
S={1\over 4g^2} \lllb (e^{-\hp} \hq e^{\hp})(e^{-\hp}\he e^\hp) -
\int_0^1 dt (e^{-t\hp}\p_t e^{t\hp})\{ (e^{-t\hp}\hq e^{t\hp}),
(e^{-t\hp}\he e^{t\hp})\}\rrrb\, ,  
\ee 
where we have divided the overall normalization by a factor
of two in order to compensate for the trace operation on the
internal matrices.
This action is invariant under the gauge transformation\cite{0001084}
\be \label{egtrsa}
\delta e^{\hp} = (\hq \wh\Omega) e^{\hp} + e^{\hp}(\he\wh\Omega') \, , 
\ee
where, as before the gauge transformation parameters $\wh\Omega$ and
$\wh\Omega'$ are
vertex operators with $(n_g,n_p)$ values $(-1,0)$ and $(-1,1)$
respectively. The internal CP indices carried by the gauge  parameters
are as follows
\be
\label{ingp}
\wh\Omega = \Omega_+ \otimes \sigma_3  + \Omega_- \otimes i\sigma_2\,, 
\ee
with a similar relation holding for $\wh\Omega'$. The GSO even $\Omega_+$ is
Grassmann odd, while the GSO odd $\Omega_-$ is Grassmann even. 
This makes the overall gauge parameters $\wh\Omega$, $\wh\Omega'$ odd
relative to $\hq$, $\he$.
The proof of gauge
invariance is formally identical to the one given in the earlier
section. The equation of motion is just \refb{eom} with hats 
on fields and operators. 
Again, the gauge parameter $\wh\Omega'$ can be used to choose the gauge
$\xi_0 \wh\Phi=0$ so we can restrict to
string states which are proportional to $\xi_0$. 

For future use, we shall now give the expansion of the action \refb{e00}
in power series in $\hp$.
Expanding the exponentials in a power series, we get,
\ben \label{eexpan}
e^{-\hp}{\cal O} e^{\hp} = \sum_{M,N=0}^\infty {1\over (M+N+1)!}
{M+N\choose M}
(-1)^M \hp^M
({\cal O}\hp) \hp^N\, ,
\een 
valid for ${\cal O}$ equal to $\hq$ or $\he$. 
Using the
cyclicity relations eq.\refb{e1aaa}, and the
identity \refb{eexpan},
we can express the
action \refb{e00} as
\be \label{e3a}
S = {1\over 2 g^2} \sum_{M,N=0}^\infty {1 \over (M+N+2)!} {M+N \choose N}
(-1)^N \lll  (\hq \hp) \hp^M (\he\hp)\hp^N \rrr\, .
\ee

As in refs.\cite{9911116,9912249} we shall find it convenient to take the
time direction
to be periodic with period 1, so that for a static configuration we can
identify the potential with the negative of the action. In this case, an
analysis analogous to that in ref.\cite{9911116} shows that the string
field
action \refb{e00} describes a D-brane with mass
\be \label{ec1}
M = {1\over 2 \pi^2 g^2}\, .
\ee
The details of this calculation have been outlined in appendix \ref{a4}.
We shall calculate the tachyon potential and attempt to show that at the
minimum it exactly cancels the mass $M$ given in eq.\refb{ec1}.

\subsection{Superstring field theory on a D-brane anti-D-brane pair} 

This system incorporates a GSO$(+)$ sector in the form of 
vertex operators that represent strings that live on
the brane or on the antibrane. With conventional Chan-Paton
indices these would use the $2\times 2$ matrices $I$ and
$\sigma_3$.  In addition there is a GSO$(-)$ sector
representing strings stretched between the brane and
antibrane.  With conventional Chan-Paton
indices these would use the $2\times 2$ matrices $\sigma_1$ and
$\sigma_2$. We will call the conventional Chan-Paton matrices
{\it external} CP matrices, to distinguish them from the
internal CP matrices used in the previous subsection.
Since we still have the complication of including two
GSO types in the string field, we will not  
dispense of the
internal CP matrices, and thus the brane-antibrane system  
will use   
{\it both} internal and external CP matrices. 
This time
we therefore write:
\be
\label{fsff}
\hp = \Phi^{(1)}_+ \otimes I\otimes I  + 
\Phi^{(2)}_+ \otimes I \otimes \sigma_3  + \Phi^{(3)}_- \otimes
\sigma_1\otimes
\sigma_1 + \Phi^{(4)}_- \otimes
\sigma_1\otimes
\sigma_2 \,,
\ee
where the first set of matrices are the internal ones and the second
set are the external ones. In computing products of fields the two
sets of matrices are defined to commute.  For the operators and 
gauge parameters we have  
\be \label{ei33}
\wh Q_B = Q_B\otimes \sigma_3\otimes I, \qquad
\wh \eta_0 = \eta_0\otimes \sigma_3\otimes I\, ,
\ee
\be
\label{ingp3}
\wh\Omega = \Omega^{(1)}_+\otimes \sigma_3 \otimes I \,+\,
\Omega^{(2)}_+\otimes \sigma_3\otimes \sigma_3
\,+\, \Omega^{(3)}_- \otimes i\sigma_2\otimes 
\sigma_1 \,+\, \Omega^{(4)}_-\otimes i\sigma_2 \otimes \sigma_2\,. 
\ee
We are still writing all fields and operators with hats, for simplicity.
The structure found earlier (eqs. \refb{e1aaa}-\refb{epr3a},
in particular) survives
when the correlators $\lll\cdots \rrr$ now include the double trace
$Tr\otimes Tr$. The action takes then the same form as in \refb{e00}
with the same normalization factor. If we restrict $\hp$ to be of the form
$\Phi\otimes I \otimes \pmatrix{1 & 0\cr 0 & 0}$, we recover the
open string
field theory action \refb{e0} on a single BPS D-brane.

As discussed elsewhere\cite{9808141}, in analyzing
the tachyon potential we can restrict ourselves to the
external CP sector $I$ in the 
GSO$(+)$ sector, and to the external
CP sector $\sigma_1$ in the 
GSO$(-)$ sector.   Thus
it is clear that there is a one to one correspondence between
the component fields of the open string field theory on the non-BPS
brane and
that on the brane anti-brane system in this restricted sector. In fact
since GSO$(-)$ fields must appear always in even numbers, the
external CP factors with their trace will simply produce an
extra factor of two for every interaction. Thus the computations
of the tachyon potential are identical. For the same value of the open
string coupling constant $g^2$, 
in the brane anti-brane
system the potential is twice as large compared to that of the non-BPS
D-brane due to the trace over the external CP factors. On the other hand
now the mass of the brane or the anti-brane
is given by $(1/2\pi^2 g^2)$, so that the total mass of the
brane-antibrane system is also twice the mass of the non-BPS D-brane.
Thus if for a non-BPS brane the potential energy at the bottom of the well
cancels the tension, then for the brane-antibrane system the potential
energy at the bottom of the well will also cancel the total tension of the
brane-antibrane system.
We will therefore
use in this paper the simpler notation required for the 
analysis of the non-BPS brane. 

\sectiono{Computation and analysis of the tachyon potential} \label{s2}

We shall be interested here 
in the phenomenon of tachyon condensation on the
non-BPS D-brane. As mentioned before, the analysis applies
also to the D-brane anti-D-brane problem.
We begin by setting up the level expansion of the
full tachyon string field relevant to the condensation. We then
discuss the expansion of the action. Finally, relegating some
computations to an appendix, we calculate the tachyon potential,
find its minimum and test the brane annihilation conjecture. 

\subsection{The tachyon string field}
 
In the present case the zero momentum tachyon
corresponds to the vertex operator $\xi ce^{-\phi}\otimes\sigma_1$. 
Let us denote by $\HH_1$ the
subset of vertex operators of ghost number 0 and picture number $0$,
created from the matter superstress tensor 
$(G_m(z), T_m(z))$,\footnote{For convenience 
of notation,
we shall denote the $k$th oscillator mode of $G_m$ and $T_m$ by $G^m_k$
and $L^m_k$ respectively.} 
and the ghost
fields $b$, $c$, $\xi$, $\eta$, $\phi$. It can be easily seen that by
restricting the string field $\wh\Phi$ to be in $\HH_1$ gives a consistent
truncation of the action, and hence we can look for a solution of the
equations of motion, representing tachyon condensation, by restricting the
string field $\wh\Phi$ to be in this subspace $\HH_1$. Thus from now on we
shall always take the string field to lie in this restricted subspace.

We now expand the
string field $\hp$ in a basis of $L_0$ eigenstates, and write the action
\refb{e3a} in terms of component fields, which are the coefficients of
expansion of the string field in this basis.
As in \cite{KS,9912249}, we shall define the level of a string field
component 
multiplying a vertex operator of
conformal weight $h$ to be $(h+{1\over 2})$, so that the tachyon field,
multiplying the vertex
operator $\xi ce^{-\phi}\otimes\sigma_1$, has level 0. We also define the
level of
a given term in the string field  action to be the sum of the levels
of the individual fields appearing in that term, 
and define a level $2n$ approximation to the action to
be the one obtained by including fields up to level $n$ and terms in the
action up to level $2n$. 
Thus for example, a level 3 approximation to the action will involve
string field components up to level (3/2). This is the approximation we
shall be using to compute the action.

\begin{table}
\begin{eqnarray*} 
\begin{array}{||c|l|c|c||} \hline
  \rule{0mm}{5mm} {L_0} & {level}
  & {GSO(+)} & {GSO(-)}
  \\ \hline 
  \rule{0mm}{6mm} -1/2 & 0 
 & {---} &  |\widetilde\Omega\rangle\\ \hline
 \rule{0mm}{6mm} 0 & 1/2 & 
c_0 \beta_{-{1\over 2}}|\widetilde\Omega\rangle
 &{---}\\ \hline
  \rule{0mm}{6mm}1/2 & 1 & {---}
& \beta_{-{1\over 2}}\gamma_{-{1\over 2}}|\widetilde\Omega\rangle\\ \hline
 \rule{0mm}{5mm} 1& 3/2& 
\{ c_{-1}\beta_{-{1\over
2}}\, ,\,b_{-1}\gamma_{-{1\over 2}},\, 
G^m_{-{3\over 2}} \} 
|\widetilde\Omega\rangle\, & {---}\\ \hline
\end{array}
\end{eqnarray*} 
\caption{Zero-momentum Lorentz scalar states of ghost
number one living in the
``small Hilbert space". Here $|\widetilde\Omega\rangle \equiv c_1
e^{-\phi(0)}|0\rangle$ is the GSO($-$) tachyon state in the conventional
minus one picture. It satisfies $L_0|\widetilde\Omega\rangle
=-{1\over 2}|\widetilde\Omega\rangle$.}
\end{table}

Using gauge invariance \refb{egtrsa} of string field theory action, we can
choose gauge conditions
\be \label{egauge} 
b_0\hp=0, \qquad \xi_0\hp=0\, . 
\ee
As in
ref.\cite{9912249}, the legitimacy of this gauge
condition  can be proved at the linearized level. We
then assume
that string field configuration under consideration is not too large, so
that such a gauge choice is also possible for the configuration under
study. Also, as discussed in ref.\cite{9912249}, the gauge choice
$b_0\hp=0$ can be made only for states with
non-zero $L_0$ eigenvalue.

We can build systematically the relevant expansion
of the string field by recalling that the 
string field $\hp$ satisfying the gauge condition $\xi_0\hp=0$ is
related to the NS string field
$\wh V$ of \cite{WITTENSFT} by the relation $\hp = \xi_0 \wh V$.
The string field $\wh V$ is built on the tachyon vacuum
$|\widetilde\Omega\rangle \equiv c_1
e^{-\phi(0)}|0\rangle$.  This vacuum state is GSO odd, it
has ghost number $+1$ and 
$L_0 = -1/2$. Being in the minus one picture, 
it is annihilated by all positively moded oscillators
$\{\gamma_r, \beta_r\}$. In addition, it is annihilated
by $b_0, L^m_{-1}$ and $G^m_{-{1\over 2}}$. 
All relevant
states of ghost number one are now obtained by acting 
with ghost number zero combinations of oscillators 
$\{b,c,\beta,\gamma, L^m, G^m\}$ on 
$|\widetilde\Omega\rangle$. The $b_0\hp=0$ gauge condition allows us to 
ignore states with a $c_0$ oscillator in them. The states one finds up to
$L_0$ eigenvalue 1 are given in Table 1. For ease of notation we have not
included the CP factor. 
Note that
we have included at $L_0=0$ a
state which is not annihilated by $b_0$.  This is the case
because having $L_0=0$ this state cannot be gauged away.

The string field we need, which uses the ``large" Hilbert space, is
obtained by acting on the states of the table with $\xi_0$.
This operation, however, does not change the dimension of the
operators. As shown in appendix \ref{a2}, 
the string field theory action in the restricted subspace $\HH_1$ has a
$Z_2$
twist symmetry
under which string field components associated with a vertex operator of
dimension $h$ carry charge $(-1)^{h+1}$ for 
even $2h$, 
and $(-1)^{h+{1\over
2}}$ for 
odd $2h$.
The tachyon vertex operator, having dimension $-{1\over 2}$, is
even under this twist transformation. Thus
we can consider a further truncation of the string field theory by
restricting the string field $\hp$  to be twist even. This, in particular,
means that the $L_0=0$ vertex operator
mentioned above is to be omitted from the
the string field. 
The same is true for the $L_0= +1/2$ state in the GSO$(-)$
sector. Therefore, in addition to the tachyon, we will
include the three scalar fields appearing in the GSO$(+)$ sector
at level $3/2$. 

In the language of vertex operators $|\wt\Omega\rangle$ is $c e^{-\phi}$,
and the
three states in table 1 at level (3/2) are 
\be \label{evert}
c \,\partial^2 c \,\partial \xi\,
e^{-2\phi} ,\quad  \eta \,, \quad G_m\, c \, e^{-\phi}\, ,
\ee
as can be seen with the help
of eq.\refb{ei1}.
We readily pass to the string field $\wh\Phi$ by acting 
the above operators with $\xi_0$, thus guaranteeing that
both gauge conditions \refb{egauge} are satisfied.
Denoting the
tachyon operator by $\wh T$ and the three other operators
by $\wh A, \wh E$ and $\wh F$ respectively, we have: 
\ben \label{e18} 
\widehat T &&= \xi\, c \,e^{-\phi}\otimes
\sigma_1 \cr \widehat A  &&= c \,\partial^2 c \,\xi\partial \xi\,
e^{-2\phi} \otimes I\cr
\widehat E  &&= \xi\,\eta \otimes I\cr
\widehat F  &&=  \xi\, G_m\, c \,e^{-\phi}
\otimes I
\een
Therefore, the
general twist even string field up to level $(3/2)$, satisfying the gauge
condition \refb{egauge}, has the following
form:\footnote{Ref.\cite{0001084} had a factor of $i$ in front of the
$t\wh T$
term. In this paper we have used slightly different set of conformal maps
$f^{(N)}_k$ in defining the string field theory action; these map the
upper half plane into the inside of the unit disk rather than outside.
With this choice, the kinetic term for the tachyon field has the
standard sign provided
there is no factor of $i$ multiplying $t\wh T$ in eq.\refb{ecc1}.} 
\be \label{ecc1}
\hp = t \,\wh T + a \,\wh A + e \,\wh E + f\, \wh F \,. 
\ee
As explained above, the tachyon vertex operator
$\widehat T$  of
$L_0 = -1/2$ is a GSO odd operator of
level zero. The operators $\widehat A,
\widehat E$ and
$\widehat F$ of $L_0 = +1$, and thus level $3/2$, are in the
GSO even sector.

\subsection{Level expansion of the string action}

We shall now substitute \refb{ecc1} into the action \refb{e3a} and keep
terms to all orders in $t$, but only up to quadratic order in $a$, $e$ and
$f$. Although the string field action contains vertices of
arbitrarily high order, it can be shown that the truncated action to any
given level only has a finite number of terms. To see this, let us first
note that for a term in the action of a given level, the number of fields
of level $>$ 0 must be
finite. Since all components of the string field other than the tachyon
$t$ has level $>$ 0, we only need to show that there cannot be
arbitrarily large
number of tachyon fields. This is easily seen by noting that the tachyon
vertex operator $\wh T$ has $-1$ unit of $\phi$ momentum. Since in order
to
get a non-zero correlation function, the total $\phi$ momentum of all the
vertex operators must add up to $-2$, it is clear that for a fixed set of
other vertex operators, we can only insert a finite number of tachyon
vertex operators in order to have a non-vanishing correlation function.

Each term in the action has one $\he$ and one $\hq$, each acting on 
a string field.
While $\he$ carries no $\phi$-momentum, the BRST operator $\hq$
can supply zero, one or two units of $\phi$-momentum (see \refb{eb4}).
The operators $\wh A$, $\wh E$ and $\wh F$ carry $-2$, 0 and $-1$ units of
$\phi$ momentum respectively.
Since the operator $E$ entering in the string field carries no $\phi$
momentum, this is the field that can appear together with 
the largest number of
tachyon fields. For example, the string action term coupling 
$E$ with four $T$'s 
is nonvanishing since the tachyons give $(-4)$ units of $\phi$-momentum
and the BRST operator can supply $(+2)$ units. Since we are going to
compute the action to level three we can have a term in the string
action with two $E$'s and four $T$'s. This is the term with the largest
possible number of fields that can contribute to level three.
This means we need the
expansion of the string action \refb{e3a} up to terms with six string
fields.
This is given by,  
\ben \label{e17p}  
S \hskip-5pt&&\hskip-10pt = {1\over 2g^2} \lllb 
\,\,{1\over 2} 
 \,(\hq \hp)\,\,(\he \hp)\, 
 + {1\over 6}
\,(\hq\hp)\, \Bigl( \hp \,(\he\hp) - (\he\hp)\, \hp\,\Bigr) \cr
&& \quad \hskip-10pt +
{1\over 24}  \,(\hq\hp) \, \Bigl( \hp^2\, (\he\hp) - 2\hp\,(\he\hp)\,\hp +
(\he\hp)\,\hp^2\, \Bigr) 
 \cr  
&& \quad\hskip-10pt +{1\over 120}
 \,(\hq\hp) \, \Bigl( \hp^3\, (\he\hp) - 3\hp^2\,(\he\hp)\,\hp
+ 3\hp\,(\he\hp)\,\hp^2 -
(\he\hp)\,\hp^3\, \Bigr) 
\cr
&& \quad\hskip-10pt +{1\over 720}\,(\hq\hp) \, 
\Bigl( \hp^4\, (\he\hp) - 4\hp^3\,(\he\hp)\,\hp
+ 6\hp^2\,(\he\hp)\,\hp^2 - 4\hp\,(\he\hp)\,\hp^2 +
(\he\hp)\,\hp^4\, \Bigr) 
\rrrb \nonumber \\
\een
Since our computation will be restricted to twist even fields in $\HH_1$,
the above result can be simplified further by 
use of \refb{tpropp} and cyclicity.
We find: 
\ben \label{e17}
S \hskip-5pt&&\hskip-10pt = {1\over 2g^2} \lllb 
\,\,{1\over 2} 
 \,(\hq \hp)\,\,(\he \hp)\, 
 + {1\over 3}
\,(\hq\hp)\,  \hp \,(\he\hp)  +
{1\over 12}  \,(\hq\hp) \, \Bigl( \hp^2\, 
(\he\hp) - \hp\,(\he\hp)\,\hp \, \Bigr) 
 \cr  
&& \qquad +{1\over 60}
 \,(\hq\hp) \, \Bigl( \hp^3\, (\he\hp) - 3\hp^2\,(\he\hp)\,\hp \Bigr) 
\cr
&& \qquad  + {1\over 360}\,(\hq\hp) \, 
\Bigl( \hp^4\, (\he\hp) - 4\hp^3\,(\he\hp)\,\hp
+ 3\hp^2\,(\he\hp)\,\hp^2  \Bigr) 
\rrrb \,. 
\een
This expansion
suffices for the present computation. All we need to 
do is to substitute \refb{ecc1}
into this
expression and evaluate the correlation functions appearing in various terms. 
The required calculations
are relatively straightforward, and
involve
correlation functions of appropriate conformal transforms of the operators
$\wh T$, $\wh A$, $\wh E$, $\wh F$, the BRST current $j_B$, and the
field $\eta$. Here we shall only state
the result; some of the details have
been discussed in appendix
\ref{a3}.\footnote{We used the symbolic manipulation program Mathematica to
carry out some of these computations.} 

\subsection{The tachyon potential}

We shall now give the result for the action and the potential by
truncating it to level 3.
Not all terms allowed by level counting
are non-vanishing. Several vanish because they fail
to satisfy $\phi$-momentum
conservation, for example, there is no $a^2t^2$ term. 

The result, with $S_k$ denoting the level $k$ terms in the action, is
\ben
g^2S_{0} \hskip-20pt&&=  {1\over 4}\,t^2 -
{1\over 2}\,t^4 \,\,,\cr \cr
g^2S_{3\over 2}\hskip-20pt &&= a\,t^2 +  
{1\over 4}\, e\,t^2 
+ {5\over 96} \sqrt{50 + 22 \sqrt{5}}\,\, 
e\,t^4\,\,,
\cr\cr 
g^2 S_{3}\hskip-20pt && = \,\,\,\, -2\,a\,e - 5\,f^2 \cr\cr
 &&\quad + \Bigl(\,{1\over \sqrt 2} - {1\over 24}\,\Bigr)\, e^2\,t^2  
- {5\over 18}  \,e^2 \,t^4
\cr\cr && \quad -\, {5\over 4} ( 4\sqrt{2}\,-1) \,f^2\,t^2  - {1 \over 12}
\,(\, 3 + 40 \sqrt{2}\,) \,ae\,t^2 + {5 \over 12}\, (\,10 \sqrt 2 - 1\,)
ef\,t^2\,.
\een
The action  to level three is given by 
$S^{(3)} = S_0 + S_{{3\over 2}} + S_3$.
Collecting the terms above, using the relation \refb{ec1}, and expressing
the various radicals as approximate decimals,  we have, 
\ben \label{er1}
V = -S &=& -M \, (2\pi^2) \bigg(0.25\, t^2 - 0.5\,t^4 + a\,t^2 + 
0.25\, e\,t^2 + 0.519\,  
e\,t^4\cr
&&\qquad\qquad 
-2\,ae - 5f^2 + 0.665\, e^2t^2 -0.278\,
 e^2 t^4 \cr
&& 
\qquad\qquad - 5.82\, f^2t^2  
\,\,- 4.96\, aet^2 \,\,+ \,\,5.476\, eft^2\,\,\bigg) \nonumber
\een
The potential has extrema at $(\pm t_0, a_0, e_0,f_0)$  
with  
\be \label{eextreme} t_0 =  0.58882, \quad a_0
=0.056363, \quad e_0 = 0.093175, \quad f_0 = 0.012603\, . 
\ee 
At these extrema,  
\be \label{epot} 
V = -0.85446 M\, .  
\ee The expected exact
answer for the value of the potential at the extrema is $-M$, so that it
can cancel the mass of the D-brane exactly. Thus we see that the level three
approximation produces 85\% of the exact answer.  Note that one finds 60\%
of the exact answer at level zero \cite{0001084}, so the approximation
scheme appears to be  
converging to the exact answer.

\begin{figure}[!ht]
\leavevmode
\begin{center}
\epsfbox{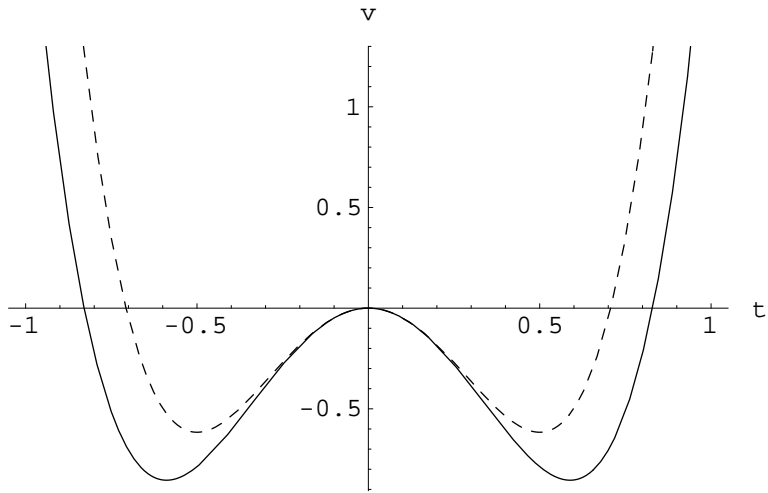}
\end{center}
\caption[]{\small The tachyon potential $v(t)=V(t)/M$ given in
\refb{epot1} (solid line).
For reference we also show
the 
zeroeth order potential (dashed line).} \label{f1}
\end{figure}

The potential  computed to this approximation (level three)
includes the fields $(a,e,f)$ only quadratically. So they can be 
integrated out exactly to find an effective potential $V(t)$ for the 
tachyon. One obtains 
\be \label{epot1}
v(t)\equiv {1\over M} \cdot V(t) = -4.93 \, t^2 \,\,{
       (1 + 4.63\, t^2 + 3.21\, t^4 - 
            9.48\, t^6 - 11.67 \,t^8)\over (1 + 
            1.16\, t^2)(1 + 2.48\, t^2)^2} \,
\ee
Several important properties are manifest from this expression.
Since the denominators never vanish, there is no singularity in
$V(t)$ to this approximation. The small tachyon instability
for $t\to 0$ is manifest. Since  $V\sim + \,t^4$, when $t$
is large,  this potential is clearly bounded below. 
It can be easily checked that only critical points are $\pm t_0$
with $t_0 = 0.58882$; 
they  are (equivalent) {\it global} minima of the
presently computed effective potential.\footnote{The critical point,
however, is not a global minima of the full multiscalar
potential $V(t,a,e,f)$. This is a reflection of the fact that some of the 
fields $a,e,f$ are auxiliary fields.} The effective
tachyon potential $v(t)$ has been displayed in fig. \ref{f1}.  

\sectiono{Tachyonic kink configuration} \label{s4}

In the last section we analysed the tachyon potential on a non-BPS D-brane
in a background independent fashion. In this section we focus on a
specific case where the non-BPS D-brane corresponds to a non-BPS D-string
of type IIA string theory wrapped on a circle of radius $R$. If $\TT_1$
denotes the tension of the non-BPS D-string, then $M=2\pi R \TT_1$. We
denote by $t$ the (1+1) dimensional tachyon field living on the non-BPS
D-string. Using the definition of $v(t)$ given in 
\refb{epot1}, we may express the potential $V(t)$ as $2\pi R \TT_1 v(t) 
= \TT_1 \int_0^{2\pi R} dx  v(t)$. From this we arrive at the conclusion
that the potential energy of the D-string is given by 
\be \label{epotden}
V(t) = \TT_1 \int_0^{2\pi R} dx v(t)\, . 
\ee 
We denote by $x^\mu$ for $\mu=0,1$ the world volume coordinates of the
D-string, and $x\equiv x^1$ is the spatial coordinte.

Since $v(t)$ given in \refb{epot1} has doubly degenerate minima
at $\pm t_0$,
we
can consider a kink solution which interpolates between these two minima.
It has been conjectured that this kink describes a BPS D0-brane of type
IIA string theory\cite{9812031,9812135}.
In this section we shall compute the mass of this kink solution, and
compare it with the
mass of a BPS D0-brane of type IIA
string theory. 

Computing the mass of the kink also requires knowledge of the kinetic term
of the tachyon.\footnote{By kinetic term we 
refer to the terms involving derivatives
of
the tachyon field $t$, including spatial derivatives.}
 We use the kinetic term obtained from the quadratic term
of the action, ignoring corrections from the higher order terms in the
presence of background $t$. There is no justification for ignoring these
corrections, and we should not expect this calculation to yield more than
an order of magnitude estimate. Using the action \refb{e00}, 
eq.\refb{ec1} relating $M=2\pi R \TT_1$ and $g^2$, and the fact that the
length of
D-string is $2\pi R$, we see that the Lagrangian contains a term
\be \label{ekin}
-{1\over 2} (2\pi^2 \TT_1) \int_0^{2\pi R} dx \p_\mu t \p^\mu t\, .
\ee
We shall now take the limit $R\to\infty$, {\it i.e.} we consider
infinitely long D-string. 
{}From eqs.\refb{epotden} and \refb{ekin} we get the following equations
of
motion for a {\it static} tachyonic configuration:
\be \label{eeqn}
2 \pi^2 \p_x^2 t = v'(t)\, .
\ee
Using the boundary condition that as $x\to\pm\infty$, $t\to\pm t_0$ and
$\p_x t\to 0$, the solution to this equation is implicitly given by:
\ben \label{esolu}
\p_x t &=& {1\over \pi} \sqrt{v(t) - v(t_0)}\, , \nonumber \\
x &=& \pi \int_{0}^{t(x)} dy {1\over \sqrt{v(y) - v(t_0)}}\, .
\een
The total energy associated with this solution (measured above the
$t=t_0$ solution), obtained by adding the
kinetic and the potential terms is given by:
\be \label{eenergy}
E =  2 \pi \TT_1 \int_{-t_0}^{t_0} dy \sqrt{v(y)- v(t_0)}\, .
\ee
Let $\TT_0$ denote the mass of a BPS D0-brane of type IIA string theory.
Then we have the relation:
\be \label{erel}
\TT_1 = \sqrt 2 {\TT_0 \over 2\pi}\, .
\ee
Using this, eq.\refb{eenergy} can be written as
\be \label{eener1}
E =  \sqrt{2} \TT_0 \int_{-t_0}^{t_0}\,dy\, \sqrt{v(y)- v(t_0)}\, . 
\ee
If we use the zeroeth order approximation\cite{0001084} for the potential
\be \label{ezer}
v(t) = 2 \pi^2 (-{1\over 4} t^2 + {1\over 2} t^4)\, ,
\ee
then \refb{eener1} can be evaluated analytically, and we get
\be \label{eana}
E = {1\over 6} \sqrt{2} \pi \TT_0\, .
\ee
This is about 74\% of the expected answer $\TT_0$.\footnote{A similar
result was independently obtained by Bergman\cite{berg}, and also by
Iqbal and Naqvi\cite{amer}.}

For the potential given  in eq.\refb{epot1}, we can calculate the right
hand side of eq.\refb{eener1} numerically. The answer is
\be \label{enum}
E = 1.03 \TT_0\, .
\ee
Considering the crude approximation that we have used, this close
agreement with the expected answer is likely to be accidental. However it
is
encouraging to note that the numerical answer is close to the expected
answer.

This analysis can be easily extended to the case of a tachyonic kink on a
non-BPS D-$p$ brane for any value of $p$.

\sectiono{Concluding remarks and open questions} \label{s5} 

There are two main points to the present paper. 
Point one:  we seem to have a consistent NS open string 
field theory \cite{9503099} in which calculations are feasible.
Point two:  the calculations performed here
with this string field theory give good direct
evidence for the tachyon condensation phenomenon
and its implications for unstable non-BPS D-branes
as well as for the D-brane anti- D-brane system. 

Let us first focus on the string field theory itself.
While the  cubic  open string field 
theory of \cite{WITTENBSFT} gives a consistent classical
theory of bosonic open strings, its extension to 
superstrings \cite{WITTENSFT} was recognized early
on to be problematic \cite{wendt}. Problems arise 
because the NS string vertex carries a picture changing
operator at the interaction point, and in testing the
associativity of the star product one induces the collision
of two picture changing operators, upon which a divergence
is encountered. It is believed that 
contact terms with infinite coefficients
must be added to the action to restore gauge invariance.
One may wonder if these complications are just irrelevant
to the problem of computing the tachyon potential.
We are not optimistic on this point.  Indeed, in this theory,
the potential of the tachyon field alone is purely quadratic.
The absence of a cubic term (because of $(-)^F$ conservation), 
and the absence of a quartic term (as the theory is cubic) 
imply that the potential for the tachyon field alone has no critical 
points. It would therefore be necessary for the 
interactions of the tachyon with the other scalars to 
generate  stabilizing terms of the right magnitude.
However, the experience in this paper, as well as that in open
bosonic string theory \cite{9912249} indicate that massive
fields rather than stabilizing the tachyon,
tend to lower the critical point that is generated
by the tachyon field alone. Given the uncertainty
in such arguments,  it would be desirable to carry
out the direct computation of the tachyon potential in
this cubic theory. Since this theory is expressed in
the ``small Hilbert space" the table given in section 3.1
lists the relevant states. Just as it was the case
in our present work,
we expect that a twist analysis will show that 
the three fields at level $3/2$ are the ones that must
be used for a lowest level nontrivial computation. 

On the other hand the WZW-like NS string field theory
used here is free of divergences and its gauge invariance
is manifest. Given that it seems now to provide a
consistent framework for dealing with the tachyon potential
in the relevant brane systems, much of our work here has
focused in the detailed setup of the action for the non-BPS
brane, as well as for the brane anti-brane systems. We 
have also given 
very explicit consideration to  the cyclicity 
and twist properties of the string action, and we
have explained in detail how to work out branch
cuts for dealing with the fractional dimension operators 
of the GSO odd sector. While this string field theory
is non-polynomial, the level expansion is workable and
the higher interactions are relatively simple to compute
since they do not involve integration over the moduli
space of Riemann surfaces; they are finite contact interactions.
In contrast to bosonic string field theory,
where gauge invariance was directly related to the
covering of the moduli space of Riemann surfaces
(see, for example, \cite{gmw} and \cite{9206084}), something
different and subtle is going on here as moduli spaces
would be covered without the help of the higher 
interactions.\footnote{Other 
approaches have been suggested
to deal with the difficulties of \cite{WITTENSFT}.
One possibility is to use string fields of non-canonical
picture number \cite{pyt,AREF2}. Another possibility is to make
the  superstring theory non-polynomial in the same way
as must be done to incorporate closed strings off-shell
\cite{openclosed}. In such approach the region of moduli space
where the collision of picture changing operators happens
is within the interaction terms, 
which could be modified
to prevent such collisions. Since the interactions in
such theory would not be of contact type the level
approximation would appear to be difficult to implement.} 

\bigskip
Turning now
to the tachyon conjectures, the results obtained
here are consistent with convergence to the expected
values. While the condensation of the 
tachyon field alone gave about 60\%
of the desired value, the first nontrivial 
correction computed here (level 3) gave about 60\%
of the remaining energy.  It should not be very
hard to use the setup of this paper to carry the 
computation to level four, and perhaps to 
automate the computation further to deal with
higher levels. Further evidence of convergence
would be desirable. It would also be of interest
to investigate further the properties of the tachyonic
kink solution representing a lower dimensional brane.
It should be noted that the convergence of the level
approximation scheme to
the answers seems slower than in the case of the
bosonic open string, where the tachyon field alone
gave about 70\% of the desired energy, and inclusion
of two additional scalars gave 95\% of the expected
answer.   
  
Of course, at a deeper level the most intriguing
questions remain those that were already apparent
in the bosonic case \cite{9912249}: (i) Is there a 
closed form solution
for the tachyon condensate? and,  (ii) what is the 
physics of the vacuum around the tachyon condensate? 
Insight into any of these two questions would open
up exciting new possibilities.

\bigskip  
\bigskip  
Acknowledgements: N.B. would like to thank Oren Bergman and Ion Vancea
for useful discussions, Caltech, Harvard University and
Massachusetts Institute of Technology for
their hospitality, and CNPq grant 300256/94-9
for partial financial support. A.S. would like to thank Caltech for
hospitality during part of this work. 
B.Z. would like to thank N. Moeller and W. Taylor for
useful discussions.

\appendix

\sectiono{Cyclicity property of string amplitudes} \label{a1} 

In this appendix we shall prove eqs.\refb{e1aa}, \refb{e1aaa}.  
Since the trace over
the Chan Paton matrices satisfy the cyclicity property without any extra
sign, we can work with the unhatted vertex operators, and prove
\refb{e1aa} and \refb{e1aaa} simultaneously. 
First we
shall prove this
for string fields 
belonging to the restricted subspace $\HH_1$, and then indicate
its
generalization for general string fields. 

\noindent
The
cyclicity properties of the conformal field theory correlation functions
are analyzed by using the property:  
\ben \label{ee1}
T\circ f^{(n)}_i\circ A &=& f^{(n)}_{i+1} \circ A
\quad \hbox{for} \quad 1\le i\le
(n-1)
\nonumber \\
T\circ f^{(n)}_n\circ A
&=& T^n\circ f_1\circ A \equiv R\circ f_1\,\circ A ,
\een
for any vertex operator A. Here  
$T(w)=e^{2\pi i/n} w$, and $R=T^n$ denotes rotation by $2\pi$. 
While the transformation $R$ acts trivially on the complex plane, 
it
must be viewed in general as the composition $T^n$ of $n$ transformations
by $T$. Thus $R$ affects the 
transformation of fields with non-integer dimension.
Since $T$
maps unit disk to itself in a one to one fashion, it corresponds to an
SL(2,R) transformation. 
Using
$SL(2,R)$ invariance of the correlation functions on the disk, we can
write
\be \label{ee2}
\langle (f^{(n)}_1\circ A_1) \cdots (f^{(n)}_{n-1}\circ A_{n-1})
(f^{(n)}_n\circ\Phi) \rangle
= \langle (f^{(n)}_2\circ A_1) \cdots (f^{(n)}_{n}\circ A_{n-1}) 
(R\circ f^{(n)}_1\circ\Phi) \rangle
\ee
In the subspace $\HH_1$, the conformal weight of $\Phi$ is integer if
$\Phi$ is Grassmann even (GSO$(+)$), and half integer if it is 
Grassman odd (GSO$(-)$). Thus
the transformation by $R$ gives a factor of 
$1$ if $\Phi$ is Grassmann even, and
$-1$ if $\Phi$ is Grassmann odd. As can be seen from 
eq.\refb{eb1}, the product of
all
the operators inside the
correlation function must be Grassmann even 
in order to get a non-vanishing
correlator. Thus we pick up a factor of 1 ($-1$) in moving the
$R\circ f^{(n)}_1\circ\Phi$
factor on the right hand side of eq.\refb{ee2} to the first place if
$\Phi$ is Grassmann even (odd). Thus the right hand
side of eq.\refb{ee2} may be written as
\be \label{ee3} 
\langle (f^{(n)}_1\circ\Phi) (f^{(n)}_2\circ A_1) \cdots (f^{(n)}_{n}\circ
A_{n-1})
\rangle
\ee
irrespective of whether $\Phi$ is Grassmann even or Grassmann odd. 

If we replace $\Phi$ by $(Q_B\Phi)$ or $(\eta_0\Phi)$, eq.\refb{ee2} still
holds, and transformation by $R$ still gives a factor of 1 ($-1)$ if
$\Phi$ is Grassmann even (Grassmann odd). 
But now, since $(Q_B\Phi)$ and $(\eta_0\Phi)$ have
statistics opposite to that of $\Phi$, we pick up a factor of $-1$ ($+1)$
in moving the $R\circ
f^{(n)}_1\circ (Q_B\Phi)$ or $R\circ 
f^{(n)}_1\circ (\eta_0\Phi)$ factor
to the first place if
$\Phi$ is Grassmann even (odd). This gives,
\ben \label{ee5}
\langle (f^{(n)}_1\circ A_1) \cdots (f^{(n)}_{n-1}\circ A_{n-1})
(f^{(n)}_n\circ(Q_B\Phi)) \rangle
&=& -\langle (f^{(n)}_1\circ(Q_B\Phi)) (f^{(n)}_2\circ A_1) \cdots
(f^{(n)}_{n}\circ
A_{n-1}) \rangle
\nonumber \\
\langle (f^{(n)}_1\circ A_1) \cdots (f^{(n)}_{n-1}\circ A_{n-1})
(f^{(n)}_n\circ(\eta_0\Phi)) \rangle
&=& -\langle (f^{(n)}_1\circ(\eta_0\Phi)) (f^{(n)}_2\circ A_1) \cdots
(f^{(n)}_{n}\circ
A_{n-1}) \rangle\nonumber \\
\een
This proves eqs.\refb{e1aa} and \refb{e1aaa}.

The cyclicity
rules derived above also hold for a general string field $\hp$ not
necessarily inside $\HH_1$, and are in fact needed for the proof of gauge
invariance of the action. The proof of these relations for a
general
D-brane system, however, requires using appropriate cyclicity axioms for the
correlation functions of a general boundary conformal field theory. In 
the present context this
axiom states that if $\Phi$ denotes a vertex operator of conformal weight
$h$, then in moving $R\circ f^{(n)}_1\circ\Phi$ from the extreme right to 
the extreme left in the right hand side of eq.\refb{ee2}, we pick up a
factor of $e^{-2\pi i h}$.  On the other hand, from eq.\refb{e3d} one can
easily check that $R=T^n$ acting on $\Phi$ gives a factor of $e^{2\pi i
h}$. Thus these two factors cancel each other, and we recover the
cyclicity rules given in the first of eqs.\refb{e1aa}, \refb{e1aaa} for a
general string
field component $\Phi$ or
$\hp$. The other two equations of \refb{e1aa}, \refb{e1aaa} can be proved
along similar
lines.

\sectiono{Twist invariance of the restricted action} \label{a2}

In this appendix we shall show that the superstring field theory action
\refb{e00}, or equivalently \refb{e3a}, has a $Z_2$ twist invariance when
we
restrict the string field $\hp$ to lie in the subspace $\HH_1$
defined in section \ref{s2}.
Using the form \refb{e3a} of the action and the
cyclicity relations given in
eq.\refb{e1aaa},
it is easy to verify that a vertex with even
number of string fields ($(M+N)$ even terms of eq.\refb{e3a}) is odd
under $Q_B\leftrightarrow\eta_0$, whereas a
vertex with odd number of string fields is even under
$Q_B\leftrightarrow\eta_0$.
Let us now
consider a typical pair of terms in the string field theory action: 
\be \label{e5}
\lll \hp^{i-1} (\hq\hp) \hp^{n-i-1} (\he\hp) \rrr
+ (-1)^{n+1} \lll \hp^{i-1} (\he\hp) \hp^{n-i-1} (\hq\hp) \rrr 
\ee

Let now  $\hp_1, \ldots \hp_n$ denote  
$n$ arbitrary components of the string field $\hp$. In the
expansion of the string action, 
the first term of \refb{e5}
will give  rise to a term of the form
\be
\label{fterm}
(I) \equiv \lll \hp_1\cdots \hp_{i-1} (\hq\hp_i) \hp_{i+1}\cdots
\hp_{n-1}(\he\hp_n) \rrr\,,
\ee
while the second term in \refb{e5} will give rise to 
a term of the form
\ben \label{e15}
(II) \equiv (-1)^{n+1} \lll \hp_{i-1}
\cdots \hp_2 \hp_1 (\he\hp_n) \hp_{n-1} \cdots \hp_{i+1} (\hq\hp_i) 
\rrr \, .
\een
In fact, when expanding the string field in arbitrary components
all terms in the action arising from \refb{e5} can be paired
just as $(I)$ and $(II)$. Note that up to a cyclic transformation,
the order of inputs in $(I)$ and $(II)$ are precisely reversed
(twisted). We will relate $(I)$ to $(II)$ up to a sign, and this
relation will enable us to derive a selection rule based on
twist.  

Let $M(z)=-z$, $\wt I(z)=(1/z)$, and $R= T^n$ denote rotation by
$2\pi$.
$R$ leaves vertex operators with integral conformal weight unchanged,
and changes the sign of the vertex operators with half-integral
conformal weight. As in the case of $f^{(N)}_k$'s, the definition of $M$
and $\wt I$ are not complete unless we specify how to choose the sign when
these transformations act on an half integral weight vertex operator. We
adopt the following convention. Acting on a primary field $\varphi$ of
weight
$h$, 
\be \label{e3e}
M\circ\varphi(z) = e^{i\pi h} \varphi(-z), \qquad
\wt I\circ\varphi(z) = (iz)^{-2h} \varphi({1\over z})\, .
\ee 
Note that since $h$ is either an integer or a half-integer, $(iz)^{-2h}$
is well defined.   
We can now verify the relations:
\ben
 \label{e7}
&&f^{(n)}_i\circ M\circ\varphi = \wt I\circ f^{(n)}_{n-i+2}\circ\varphi 
\quad \hbox{for} \quad n\ge i\ge  2\, \cr  
&& f^{(n)}_1\circ M\circ\varphi = \wt I\circ R\circ f^{(n)}_1\circ\varphi,
\een
where the second equation is clearly the natural generalization
of the first once we note that $f^{(n)}_{n+1}$ 
is identified with $R\circ f^{(n)}_1$.
Since secondary vertex operators are obtained from 
products of derivatives of primary vertex
operators, these relations also hold if we replace
$\varphi$ by a secondary vertex operator.

\medskip 
We now  consider $(I)$ which explicitly reads
\ben \label{e6}
(I) &=& 
Tr\langle
(f^{(n)}_1\circ\hp_1)\cdots (f^{(n)}_{i-1}\circ\hp_{i-1})(f^{(n)}_i\circ 
(\hq\hp_i))(f^{(n)}_{i+1}\circ \hp_{i+1})\nonumber \\
&& \qquad \qquad \cdots 
(f^{(n)}_{n-1}\circ\hp_{n-1})
(f^{(n)}_n\circ(\he\hp_n)) \rangle\, , 
\een
where for simplicity we have omitted the zeroes from the arguments of
$\hp_i$. 
Since $M$ preserves the origin of the coordinate system, using
\refb{e3e},
we can replace each vertex operator $\hp_i(0)$ in the 
above 
correlator by $e^{-i\pi h_i} M\circ\hp_i(0)$, where $h_i$ is the
conformal dimension of $\hp_i$.
We then use 
\refb{e7} to bring 
\refb{e6} into the form: 
\ben \label{e8}
 && (I) =(-1)^{\sum h_i} 
Tr\langle (\wt I\circ R\circ 
f^{(n)}_1\circ\hp_1) (\wt I \circ f^{(n)}_n \circ \hp_2) \cdots
(\wt I \circ f^{(n)}_{n-i+3}\circ\hp_{i-1}) \nonumber \\
&& 
\hskip-18pt (\wt I\circ f^{(n)}_{n-i+2}\circ 
(\hq\hp_i)) (\wt I\circ f^{(n)}_{n-i+1}\circ \hp_{i+1})\cdots
(\wt I\circ f^{(n)}_{3}\circ\hp_{n-1})
(\wt I\circ f^{(n)}_2\circ(\he\hp_n))\rangle\, . 
\een
We now use the following results:
\begin{itemize}
\item In the restricted sector in which we are working, the SL(2,C)
transformation $\wt I$ is a symmetry of the correlation functions.
Thus we
can remove all factors of $\wt I$ from eq.\refb{e8}.
\item Acting on $\hp_1$, $R$ gives a factor of $(-1)^{2h_1}$.
\item If we reverse the ordering of $f^{(n)}_n\circ\hp_2\ldots
f^{(n)}_{n-i+2}\circ (\hq\hp_i) \ldots
f^{(n)}_2\circ(\he\hp_n)$ in eq.\refb{e8}, then we pick up a factor of
$(-1)(-1)^{n_o'(n_o'-1)/2}$, where $n_o'$ is the number of odd string
fields in the set $\hp_2, \ldots \hp_n$. The first minus 
sign in this expression comes from passing $\wh Q_B$ through 
$\wh \eta_0$; the other factor
comes from passing the odd components of the string field through each
other. Note that due to the internal CP matrices 
there is no extra sign in passing 
$\wh Q_B$ or $\wh\eta_0$ through
$\hp_j$, irrespective of whether $\hp_j$ is Grassmann even or odd. 
\end{itemize}
Thus \refb{e8} can be
written as
\ben \label{e9}
&&(I) = (-1)^{\sum h_i} (-1)^{1+2h_1+n_o'(n_o'-1)/2}\,
Tr\langle (f^{(n)}_1\circ\hp_1)  
(f^{(n)}_2\circ(\he\hp_n)) (f^{(n)}_{3}\circ\hp_{n-1})\nonumber \\
&& \qquad  \cdots
(f^{(n)}_{n-i+1}\circ \hp_{i+1}) (f^{(n)}_{n-i+2}\circ 
(\hq\hp_i)) 
(f^{(n)}_{n-i+3}\circ\hp_{i-1}) \cdots (f^{(n)}_n \circ \hp_2)
\rangle \nonumber \\
& =  & (-1)^{\sum h_i} (-1)^{1+2h_1+n_o'(n_o'-1)/2}
\lll \hp_1 (\he\hp_n) \hp_{n-1} \cdots \hp_{i+1} (\hq\hp_i) \hp_{i-1}
\cdots \hp_2 \rrr  
\, . 
\een

Let $n_e$ and $n_o$ denote the total number of even and odd fields in the
set $\hp_1,\ldots \hp_n$. Since $n_o$ is always even, we may write
$n_o=2m$ for some integer $m$. We now analyse two cases
separately.
\begin{enumerate}
\item $\hp_1$ odd. In this case $n_o'=2m-1$ and $(-1)^{2h_1}=-1$. Thus:
\be \label{e12}
(-1)^{2h_1} (-1)^{n_o'(n_o'-1)/2} = (-1)^{1+(m-1)(2m-1)}=(-1)^m\, ,
\ee
for integer $m$. 
\item $\hp_1$ is even. In this case $n_o'=2m$, $(-1)^{2h_1}=1$,
and we have
\be \label{e13}
(-1)^{2h_1} (-1)^{n_o'(n_o'-1)/2} = (-1)^{m(2m-1)}=(-1)^m\, .
\ee
\end{enumerate}
Thus in both cases $(-1)^{2h_1}(-1)^{n_o'(n_o'-1)/2}=(-1)^m$. Using
eqs.\refb{e12}-\refb{e13}, and the cyclicity property \refb{e1aaa}, we can
finally express the right hand side of \refb{e9}
as
\be \label{e14}
(I) = (-1)^{{n_o\over 2}+1}(-1)^{\sum h_i}
\lll \hp_{i-1}
\cdots \hp_2 \hp_1 (\he\hp_n) \hp_{n-1} \cdots \hp_{i+1} (\hq\hp_i) 
\rrr \, .
\ee
We now recognize that the operators inside the correlator
are ordered just as in \refb{e15}. Since the total contribution to 
the action is given by the addition of $(I)$ and $(II)$, 
combining \refb{e14} and
\refb{e15} we see that we get a non-zero contribution only if
\be \label{e15a}
(-1)^{n+{n_o\over 2}} (-1)^{\sum_i h_i} =1\, .
\ee
Since $n_o$ is always even, we have $(-1)^n=(-1)^{n_o+n_e}=(-1)^{n_e}$.
Thus we may rewrite \refb{e15a} as
\be \label{e16}
(-1)^{\sum_{even} (h_i+1)}(-1)^{\sum_{odd} (h_i+{1\over 2})} = 1\, .
\ee
This can be interpreted by saying that the action has a $Z_2$ twist
invariance under which even
fields carry twist charge $(-1)^{h+1}$ and odd fields carry twist charge
$(-1)^{h+{1\over 2}}$. This means, in particular, that the tachyon $\wh T$,
being Grassmann odd and of dimension $-1/2$ has twist charge $+1$. In the
computation of the tachyon potential we can therefore restrict
${\cal H}_1$ to twist even fields. 

The results of this appendix  can be used to relate terms  
in the string action. It follows from our discussion above that
for a vertex involving $n$ string 
fields in $\HH_1$: 
\be
\label{tprop}
\lll\hp_1 \cdots (\hq\hp_k) \cdots 
(\he\hp_l) \cdots \hp_n \rrr = (-)^{n+1} \Bigl(
\prod_{i=1}^n \Omega_i \Bigr)
\lll \hp_n  \cdots (\he\hp_l) \cdots (\hq\hp_k) \cdots \hp_1 
\rrr
\ee
where $\Omega_i$ is the twist eigenvalue of $\hp_i$.
When we restrict to twist even 
fields in ${\cal H}_1$ the
above equation 
is even simpler:
\be
\label{tpropp}
\lll\hp_1 \cdots (\hq\hp_k) \cdots 
(\he\hp_l) \cdots \hp_n \rrr = (-)^{n+1} 
\lll \hp_n  \cdots (\he\hp_l) \cdots (\hq\hp_k) \cdots \hp_1
\rrr \, . 
\ee

\sectiono{Mass  of the D-brane} \label{a4}

In this appendix we shall show that 
the mass of the D-brane, whose world volume theory is given by
the
action
\refb{e00}, is given by $(2\pi^2 g^2)^{-1}$. 
The strategy that we shall be following is as follows.
As in ref.\cite{9911116}, we
assume that there is a set of (at least
one) non-compact flat directions transverse to the D-brane; we shall
denote
these coordinates
by
$x^i$. Then the open string modes living on the D-brane will include the
location of the D-brane along the directions $x^i$. Let $Y^i$ denote the
coordinate of the D-brane along $x^i$. The string field theory
action contains a term
proportional to $(\p_t Y^i)^2$, where $\p_t$ denotes time derivative.
The coefficient of the $(\p_t
Y^i)^2$ can be identified as half of the D-brane mass.

Let $X^i$ be the free world-volume scalar field associated with the
coordinate 
$x^i$, and $\psi^i$, $\wt\psi^i$
its left- and right-handed
supersymmetric partners. 
We denote by $x^0\equiv t$ the time
coordinate, $X^0$ the
corresponding world-volume scalar field, and $k_0$ the quantum number
labelling momentum conjugate to $X^0$. If we write $X^\mu=X^\mu_L+X^\mu_R$
with $L$ and $R$ denoting left and right-moving components, then, 
\be \label{exnorm}
\p X^\mu_L(z) \p X^\nu_L(w) \simeq - { \eta^{\mu\nu} \over
2 (z-w)^2}\, \, ,
\quad
\psi^\mu(z) \psi^\nu(w) \simeq {\eta^{\mu\nu}\over 2(z-w)}\, ,
\ee
with $\eta^{\mu\nu}=diag(-1,1,\ldots,1)$. 
With this normalization,
\be \label{enorm2}
T_m = - \p X_L \cdot\p X_L - \psi \cdot
\p
\psi + \cdots,
\quad G_m = 2i \,\psi \cdot \p X_L + \cdots \, .
\ee
There is a similar set of relations
for the right-moving (anti-holomorphic) fields.

Since the time direction has been
taken to be periodic with period 1, $k_0$ is quantized in units of $2\pi$.
Let us now consider the following term in the expansion of the
string field $\hp$;
\be \label{ed1}
\hp =
\sum_{k_0} \phi^i(k_0) \sqrt{2} \xi c e^{-\phi} \psi^i e^{i k_0X^0}\otimes
I + \cdots\, .
\ee
The $\sqrt{2}$ factor in this expansion has been included to compensate
for the factor of (1/2) in the operator product of $\psi^i$ with itself.
Although $X^0=X^0_L+X^0_R$, using the Neumann boundary condition
$X_L=X_R$ at the boundary we can replace $e^{ik_0X^0}$ by
$e^{2ik_0X^0_L}$. This facilitates computation of various correlation
functions. In particular, using eqs.\refb{exnorm}, \refb{enorm2} we see
that this vertex operator has $L^m_0$ eigenvalue equal to
${1\over 2}-(k_0)^2$, where $L^m_k$ denotes the $k$th 
mode of the matter
Virasoro generator.

 We shall now examine
the quadratic term in the
action involving the mode $\phi^i(k_0)$.
Only the $c T_m$ term of the
BRST current $j_B$ contributes to the $k_0$ dependent part of the
quadratic
term involving this mode, and
the result is given by
\be \label{e3.5}
{1\over 2 g^2} \sum_{k_0} (k_0)^2
\phi^i(k_0)\phi^i(-k_0)\, ,
\ee
in the $\alpha'=1$ unit. If $\chi^i(t)\equiv \sum_{k_0} e^{i k_0 t}
\phi^i(k_0)$ denotes the Fourier transform of
$\phi^i(k_0)$, then the above action can be rewritten as
\be \label{e3.6}
{1\over 2 g^2} \int dt \p_t \chi^i \p_t \chi^i\, ,
\ee
where $t\equiv x^0$ denotes the time variable conjugate to $k_0$. 

Up to an
overall
normalization factor, 
$\chi^i$ has the interpretation of the location $Y^i$ of the D-brane in
the
$x^i$
direction. We shall now determine the normalization factor between
$\chi^i$ and $Y^i$. For this, instead
of taking a single D-brane, let us take a pair of identical D-branes,
separated by a distance $b^i$ along the $X^i$ direction. Then each state
in the open string Hilbert space carries a $2\times 2$ Chan Paton factor,
besides the usual CP factor carried by a single non-BPS D-brane; we shall
call these external CP factors.
States with off diagonal external CP factors, representing open
strings
stretched between the two branes, are forced to carry an
amount of winding charge $b^i$ along $X^i$. For $\alpha'=1$, {\it i.e.}
string tension$=(2\pi)^{-1}$, the classical contribution to the
mass of these open string
states due to the tension of the string is equal to $|\vec b|/(2\pi)$. If
we now move one of the
branes by an amount $Y^i$ along $X^i$, the change in the (mass)$^2$ of the
open string with Chan Paton factors
$\pmatrix{0 & 1\cr 0 & 0}$ and $\pmatrix{0 & 0\cr 1 & 0}$ should be
given by:
\be \label{ecpone}
{1\over (2\pi)^2} \{(\vec b+\vec Y)^2 -\vec b^2\} = {1\over 2\pi^2} \vec
b\cdot \vec Y + O(\vec Y^2)\, .
\ee

On the other hand, since $\chi^i$ denotes the mode which translates the
brane, 
moving {\it one of the branes} along $X^i$ will correspond to
switching on a constant $\chi^i$. This is represented by a string field
background 
\be \label{ecptwo}
\sqrt{2} \chi^i \xi c e^{-\phi} \psi^i \otimes I \otimes \pmatrix{1 & 0\cr
0 & 0}\, .
\ee
We shall now explicitly use the string field theory action \refb{e00} to
calculate the
change of the (mass)$^2$ of states with Chan Paton factors $\pmatrix{0 &
1\cr 0
& 0}$ and $\pmatrix{0 & 0\cr 1 & 0}$ due to the presence of this
background string field, and compare with eq.\refb{ecpone}. For this we
note that the vertex operator for the lowest mass open
string with internal CP factor $I$ and
external CP factors $\pmatrix{0 & 1\cr 0 & 0}$ and $\pmatrix{ 0 & 0\cr 1 &
0}$ are given by, respectively,
\ben \label{ey1}
&& \xi c e^{-\phi} (\vec\epsilon\cdot \vec\psi)e^{i {b^i\over
2\pi}(X^i_L-X^i_R)}
e^{2ik_0X^0_L}\otimes I\otimes
\pmatrix{0 & 1\cr 0 & 0}\, , \quad \hbox{and} \nonumber \\
&& \xi c e^{-\phi} (\vec\epsilon\cdot \vec\psi) e^{-i {b^i\over
2\pi}(X^i_L-X^i_R)}
e^{2ik_0X^0_L}\otimes I\otimes
\pmatrix{0 & 0\cr 1 & 0}\, ,
\een
where $\vec\epsilon$ is a polarization vector. Using
Dirichlet boundary condition on $X^i$, we can write
$X^i_L-X^i_R=2 X^i_L$. Requiring BRST invariance of these vertex operators
gives,
\be \label{ey2}
\vec\epsilon\cdot \vec b =0, \qquad (k_0)^2={\vec b^2\over (2\pi)^2}\, .
\ee
Thus they represent states of mass $|\vec b|/(2\pi)$. We shall normalize
$\vec\epsilon$ such that 
\be \label{ey3}
|\vec\epsilon|^2=2\, .
\ee

Let us now consider the following expansion of the string field
\be \label{ey4}
\hp = \chi^i\wh P^i + \sum_{k_0} (u(k_0) \wh U(k_0) + u^*(k_0) \wh V(k_0))
+
\ldots
\ee
where 
\be \label{ey5}
\wh P^i = \sqrt{2} \xi c e^{-\phi} \psi^i 
\otimes I \otimes \pmatrix{1 & 0\cr
0 & 0}\,  ,
\ee
\be \label{ey6}
\wh U(k_0) = \xi c e^{-\phi} (\vec\epsilon\cdot \vec\psi)e^{2i {b^i\over
2\pi}X^i_L}
e^{2ik_0X^0_L}\otimes I\otimes
\pmatrix{0 & 1\cr 0 & 0}\, ,
\ee
\be \label{ey7}
\wh V(k_0) = \xi c e^{-\phi} (\vec\epsilon\cdot \vec\psi) e^{-2 i
{b^i\over
2\pi} X^i_L}
e^{2ik_0X^0_L}\otimes I\otimes
\pmatrix{0 & 0\cr 1 & 0}\, .
\ee
$\chi^i$, $u(k_0)$ and $u^*(k_0)$ are specific components of the string
field. We can now evaluate the string field theory action as a function of
these fields. We shall be interested in the quadratic term involving $u$, 
$u^*$,
as well as the $\chi^iuu^*$ coupling. The quadratic term is given by
\be \label{ey8}
{1\over g^2} \sum_{k_0} u^*(-k_0) u(k_0) (k_0^2 - {\vec b^2\over
4\pi^2})\, .
\ee
The computation of the $\chi^iuu^*$ coupling can be simplified if we 
work on-shell at $k_0^2=\vec b^2 / (2\pi)^2$. (This  suffices for
computing the shift in mass$^2$ of the state to order $\chi^i$.) 
We now
note that: \begin{itemize}
\item Using the three point vertex
$(12g^2)^{-1}(\lll(\hq\hp)\hp(\he\hp)\rrr-\lll\hp(\hq\hp)(\he\hp)\rrr)$ we
get
twelve terms contributing to the 
$\chi^iuu^*$ coupling. Half of these terms vanish due to the trace
identity:
\be \label{ey9}
Tr\bigg(\pmatrix{0 & 0\cr 1 & 0} \pmatrix{0 & 1\cr 0 & 0} \pmatrix{1 &
0\cr 0 & 0}\bigg)
= 0\, .
\ee
The other cyclic ordering of these matrices produce a non-zero answer
(equal to unity)
for this trace. 
\item Each of the vertex operators $\wh P^i$, $\wh U$ and $\wh V$ is
annihilated by $\wh Q_B\wh\eta_0$ if $k_0^2=\vec b^2/(2\pi)^2$. Using this
result we can manipulate each of
the remaining six terms so that $\hq$ acts on $\wh P^i$, and $\he$ acts
on $\wh U$. Finally, using the cyclicity relations \refb{e1aaa} we can
show
that each of these six terms gives identical result proportional to
$\lll(\hq\wh P^i)(\he\wh U)\wh V\rrr$.
\end{itemize}
After performing the trace over CP factors, and restricting to only on-shell
components of $u$ and $u^*$, we may express the $\chi^iuu^*$
term in the action as:
\be \label{ey10}
-{1\over g^2}
{\sum_{k_0}}' \chi^i u^*(-k_0) u(k_0) \langle
f_1\circ(Q_BP^i) f_2\circ(\eta_0 U(k_0)) f_3\circ V(-k_0)
\rangle\, ,
\ee
where $\sum'$ denotes sum over on-shell momenta $k_0=\pm |\vec b|/(2\pi)$.
This correlation function is easily evaluated and the result is
\be \label{ey11}
{1 \over g^2} {1\over \sqrt 2\pi} \vec b\cdot \vec \chi {\sum_{k_0}}'
u^*(-k_0) u(k_0)\, .
\ee
Combining this with eq.\refb{ey8} we see that the shift in the mass$^2$ of
the $u$, $u^*$ field due to the presence of $\chi^i$ background is given
by
\be \label{ecp3}
-{1\over \sqrt 2\pi} \vec b\cdot \vec \chi + O(\vec \chi^2)\, .
\ee
Comparing eqs.\refb{ecpone} and \refb{ecp3}
we get
\be \label{ecp4}
\chi^i=-{Y^i\over \sqrt 2\pi}\, .
\ee
Once we have determined the relative normalization between $\chi^i$ and
$Y^i$, we can return to the system containing a single
brane.\footnote{This
can be done, for example, by moving the other brane infinite distance away
by taking the limit $|\vec b|\to\infty$.} Substituting eq.\refb{ecp4}
into eq.\refb{e3.6}, we get,
\be \label{ecp5}
(4\pi^2 g^2)^{-1}\int dt \p_t Y^i\p_t Y^i\, .
\ee
This contribution to
the D-brane world-volume action can be
interpreted as due to the kinetic energy associated with
the collective motion of the D-brane in the non-compact transverse
directions. This allows
us to identify the D-brane mass as
\be \label{e3.7}
M=(2\pi^2 g^2)^{-1}\, .
\ee

\sectiono{Details on the calculation of the tachyon potential} \label{a3}

We first consider some of the ingredients of the calculation, then
do a particular example in 
detail.  
First of all, computation of $\lll~\rrr$ involving the various vertex
operators $
T, A, \ldots$ requires knowledge of $f\circ  T(0)$, $f\circ
 A(0)$ etc., for a
conformal map $f$. If $f(0)=w$, then we have the following relations:
\ben \label{e18xx} 
f\circ  T(0) &=& (f'(0))^{-{1\over 2}}\,  T(w) \cr 
f\circ  A(0)  &=& f'(0) \Bigl( A(w) -{f''(0)\over (f'(0))^2} c
\,\partial c \,\xi\partial \xi\,
e^{-2\phi}(w) \Bigr) \cr
f\circ  E(0)  &=& f'(0) \Bigl(  E(w) -{f''(0)\over 2 (f'(0))^2}
\Bigr) \cr
f\circ F(0)  &=&  f'(0)  F(w)
\een
Since the action involves
$Q_B$ and $\eta_0$ acting on string fields, 
we need to evaluate
those
on $T,A,E$ and $F$ and the result of conformal transform of these
operators. However the analysis can be simplified by noting that 
\be \label{esimpli}
f\circ(\OO A) = \OO(f\circ A)\, ,
\ee
where $\OO$ can be either $Q_B$ or $\eta_0$. This is due to the fact that
the BRST current $j_B$ and $\eta$ are dimension 1 primary fields. Thus
for example, in
calculating correlation function involving $f\circ(Q_B A(0))$ we need to
calculate the correlation function involving $j_B(w)f\circ A(0)$ and pick
up the residue of the pole at $w=f(0)$. A similar procedure holds
for
$f\circ(\eta_0 A(0))$.

These relations, together with eq.\refb{eb6}, and the identity
\ben \label{ecorrln}
&& \langle \,
\prod_{i=1}^{n+1} \xi(x_i)   
\prod_{j=1}^{n} \eta(y_j)  
\prod_{k=1}^m
b(u_k) \prod_{l=1}^{m+3} c(v_l) \prod_{s=1}^p e^{q_s\phi(z_s)}\rangle 
\nonumber
\\
= &&\hskip-10pt  -\prod_{i<i'} (x_i - x_{i'}) \prod_{j<j'} (y_j - y_{j'})
\prod_{i,j} (x_i
- y_j)^{-1} \prod_{k<k'} (u_k - u_{k'}) \prod_{l<l'} (v_l - v_{l'})
\prod_{k,l} (u_k - v_l)^{-1} \nonumber \\
&& \times \prod_{s<s'} (z_s - z_{s'})^{- q_s q_{s'}}\, , 
\een
allows us to compute the relevant terms 
which appear in the computation of 
the tachyon potential. Eq.\refb{ecorrln} follows from the normalization
convention \refb{eb1}, and the operator products \refb{eb3}. 

In evaluating correlation functions involving the operator $E$, we need to
exercise special care, as it
involves product of $\xi$ and $\eta$ at the same point. This has to be
interpreted as:
\be \label{exieta}
\xi\eta(w) = \lim_{z\to w} \bigg(\xi(z) \eta(w) - {1\over z-w}\bigg)\, .
\ee

Let us give as an example the computation of the quartic
term in the tachyon potential.  From the expansion
of the action \refb{e17}, focusing on the terms with
four string fields,
we find: 
\ben
\label{compu4}
g^2 S\, \Bigl|_{t^4} &&= -{t^4\over 24} \Bigl\{ \lll \,(\hq\wh T )\,
\wh T\,(\he\wh T)\,\wh T\rrr - \lll\,(\hq\wh T) \,
\wh T\,\wh T(\he\wh T)\,
  \rrr \Bigr\}\,,\cr
&&= -{t^4\over 12} \Bigl\{ \lll \,(Q_B T )\,
 T\,(\eta_0 T)\, T\rrr + \lll\,(Q_B T) \,
 T\, T(\eta_0 T)\,
  \rrr \Bigr\}\,.
\een
In the second step we evaluated the trace over the
internal CP matrices. We therefore have two correlators
to compute.  Using the fact that $T$ correspond to a dimension $-(1/2)$
primary field, and that both $j_B(w)$ and $\eta(w)$ have only single
poles near a $T$, the first correlator
in the above equation can be written as:
\ben
C(f_1,f_2,f_3,f_4)  &\equiv& \langle \,f_1\circ (Q_B T(0) )\,
 f_2\circ T(0) \,f_3 \circ (\eta_0 T(0))\, f_4\circ T(0)\rangle\cr\cr
&=& \lim_{y_1\to w_1}\lim_{y_2\to w_3} (y_1-w_1) (y_2-w_3)
{\langle
\,  j_B(y_1) T(w_1)\,
  T(w_2) \, \eta(y_2) T(w_3)\,  T(w_4) \rangle
\over 
(f_1')^{1\over 2}(f_2')^{1\over 2}
(f_3')^{1\over 2}(f_4')^{1\over 2}}\, , \nonumber \\
\een 
where $w_i=f_i(0)$. We have, for simplicity of notation, defined
$f_i\equiv f^{(4)}_i$. This correlation function can be easily evaluated,
and the answer is
\be \label{eans}
C(f_1,f_2,f_3,f_4) = {w_{13}w_{24}\over
(f_1')^{1\over 2}(f_2')^{1\over 2}
(f_3')^{1\over 2}(f_4')^{1\over 2}}\, ,
\ee
where $w_{ij}=(w_i-w_j)$.
We now recognize that
the second correlator in \refb{compu4} is simply
$C(f_1,f_2,f_4,f_3)$ with no extra sign factor because 
the last two vertex operators do not induce a sign
factor when they are transposed. We can therefore
write the complete answer as 
\be
\label{cans}
g^2 S\,\Bigl|_{t^4}= -\, {w_{13}w_{24}+ w_{14}w_{23}\over 12\,
(f_1')^{1\over 2}(f_2')^{1\over 2}
(f_3')^{1\over 2}(f_4')^{1\over 2}} \, t^4\, .
\ee
This off-shell amplitude is PSL(2,C) invariant\footnote{See 
\cite{9409015} for a Riemann surface interpretation of invariant
off-shell  amplitudes.}. Indeed letting 
\be
w \to {aw+ b\over cw+d}\, , \quad ad-bc=1,
\ee
we readily find that
\be
w_{ij} \to {w_{ij}\over (cw_i + d)(cw_j + d)  }\,, \qquad
f_i' \to {f_i' \over (cw_i + d)^2}
\ee
and therefore we get PSL(2,C) invariance if we choose the 
branch\footnote{In the chosen SL(2,C) transformation
there is a sign ambiguity in
which all coefficients $a,b,c,d$ of the transformation are 
changed in sign. Since
this
transformation must be used for an even number of punctures, this is not 
a problem. }
\be
\label{trh} 
(f_i')^{1/2} \to {(f_i')^{1/2} \over cw_i + d}\,.
\ee
We evaluate now the term.  Our first choice of coordinates 
is that of the unit disk, described in detail in section 2.1.
The prescription for dealing with the square roots there 
(\refb{e3d})
is used to find 
\be
\label{cans1}
g^2 S\,\Bigl|_{t^4}= -\, {2(2i) +  (1+i)^2\over 12\cdot
e^{i\pi/2}}\, t^4 = -{1\over 2}\, t^4
\ee
which is the result obtained in \cite{0001084}. 

We shall
now do the computation in the upper half plane (UHP) using the maps
$g^{(n)}_k$, related to $f^{(n)}_k$ by an SL(2,C) transformation which
maps the disk to UHP. But before
we proceed, we need to derive the analog of eq.\refb{e3d} for half integer
$h$,
{\it i.e.} the presription for choosing the sign of
$(g^{(n)\prime}_k(0))^{1\over 2}$ appearing in the conformal transform of
half-integer weight fields. This will be done by starting with the
presciption \refb{e3d} and then using prescription \refb{trh} for an
appropriate SL(2,C) transformation relating $f^{(n)}_k$ to $g^{(n)}_k$.
First note that for fixed $n$ and $k$, 
$f^{(n)}_k(z)$ moves anti-clockwise
along the boundary of the unit disk as $z$ moves along the positively
oriented real line.\footnote{This
is related to the fact that the canonical half disks representing the
strings are all mapped analytically into the interior of the 
disk and  the boundary of the canonical half-disks is oriented
in the direction of increasing real values.}  In addition, since
the map from the disk to the UHP takes the anti-clockwise oriented
boundary of the disk to the  
positively oriented real line,  it is clear that the
$g^{(n)\prime}_k$'s map to positive real values at the punctures on the
real line. 
In computing $(g^{(n)\prime}_k)^{1\over 2}$ we
have 
a sign ambiguity. We shall now show that {\it if the conformal map
relating $f^{(n)}_k$ to $g^{(n)}_k$ is such that the points $g^{(n)}_1(0),
\ldots g^{(n)}_n(0)$ are ordered from the left to the right on the real
axis, then we should choose the positive sign for all the
$(g^{(n)\prime}_k(0))^{1\over 2}$.}

We prove this as follows. As a first step it is convenient
to rotate the punctures on the disk to a new position.  
For this, we define:
\be \label{eprr1}
\wt f^{(n)}_k(z) = e^{{2\pi i\over n} (-{n\over 2}+1 - \epsilon)}
f^{(n)}_k(z)
=e^{{2\pi i\over n} ( k - {n\over 2} - \epsilon)}\bigg({1 + iz \over 1 -
iz}\bigg)^{2\over n}\, ,
\ee
where $\epsilon$ is a small positive number; in fact any $0<\epsilon<1$
will do. 
In this case
\be \label{eprr2}
(\wt f^{(n)\prime}_k(0))^{1\over 2} = e^{{i\pi\over n} (-{n\over 2}+1 -
\epsilon)}
(f^{(n)\prime}_k(0))^{1\over 2} = \bigg|\bigg({4\over n}\bigg)^{1\over 2}
\bigg|\,  
e^{{i\pi\over n}(k -{n\over 4}
-\epsilon)}\, .
\ee
Next we define
\be \label{eprr3}
g^{(n)}_k(z) = F(\wt f^{(n)}_k(z))\, ,
\ee
with 
\be \label{eprr4}
F(u)=i\,{1-u\over 1+u} \equiv {a u + b\over cu + d}\, , 
\ee
where we use our freedom to fix the signs of $a,b,c,d$ to write: 
\be \label{rprr5}
\pmatrix{a & b\cr c & d} = \pmatrix{{1\over \sqrt 2} e^{-{i\pi\over 4}} & 
- {1\over \sqrt 2} e^{-{i\pi\over 4}} \cr
{1\over \sqrt 2} e^{{i\pi\over 4}} & {1\over \sqrt 2} e^{{i\pi\over 4}}},
\qquad ad - bc=1\, .
\ee
$F$ describes an SL(2,C) map from the disk to the UHP.
With this,
\be \label{eprr6}
g^{(n)}_k(0)=\tan \Big({\pi\over n}(k - {n\over 2} - \epsilon)\Big)\, .
\ee
For $k=1,\ldots n$, the $g^{(n)}_k(0)$'s given 
above are arranged from left to
right on the real axis. We also have
\be \label{eprr7}
(g^{(n)\prime}_k(0))^{1\over 2} = 
(c \wt f^{(n)}_k(0) + d)^{-1} 
(\wt f^{(n)\prime}_k(0))^{1\over 2} = {1\over \sqrt 2} 
\bigg|\bigg({4\over n}\bigg)^{1\over 2}
\bigg|\, 
\sec\Big({\pi\over n}(k
-{n\over 2} -\epsilon)\Big)\, .
\ee
This is manifestly positive for $1\le k\le n$.

This gives one set of $g^{(n)}_k$'s for which the square root rules stated
above hold, but we need to show that this holds for any other set of
functions $\wt g^{(n)}_k(z)$, related to $g^{(n)}_k(z)$ by an SL(2,R)
transformation. For this, 
let us consider another set of functions $\wt g^{(n)}_k$'s related to
the $g^{(n)}_k$'s via an SL(2,R) transformation $\pmatrix{p & q\cr r & 
s}$ with the property that
$\wt g^{(n)}_1(0),\ldots \wt g^{(n)}_n(0)$ are arranged from the left to
the right on the real axis.  In that case, 
if $v_k=g^{(n)}_k(0)$, then for
$k>l$,
\be \label{eqrr1}
v_k> v_l, \qquad {p v_k + q\over r v_k + s} - {p v_l + q\over r v_l + s} =
{(v_k - v_l)\over (r v_k + s) (r v_l + s)} > 0\, .
\ee
Thus
\be \label{eqrr2}
(r v_k + s) (r v_l+s)> 0\, .
\ee
This shows that $(r v_k+s)$ has the same sign for all $k$. Using the
freedom of changing the sign of $p,q,r,s$, we can take $(r v_k+s)$ to
be positive. Then
\be \label{eqr0}
(\wt g^{(n)\prime}_k(0))^{1\over 2} = 
(r v_k + s)^{-1} 
(g^{(n)\prime}_k(0))^{1\over
2} > 0\, .
\ee
This proves the desired result.

Let us now get back to the computation of \refb{cans} using maps to UHP.
For this we map
the disk, punctured at $1,i, -1,-i$, into the UHP with
the real boundary punctured at $-4,-1,0,2$.  
These are
particularly nice points that give coordinates without
radicals: 
\ben
g^{(4)}_1(z) &&= - 4 + 6\,z
\,\,-9\,z^2+\cdots
\cr g^{(4)}_2(z) &&= -1 +
{\textstyle{3\over 4}} \, z - {\textstyle{3\over 16}}  \,
z^2   +
\cdots \cr
g^{(4)}_3(z) &&=\,\,\, 0  + \,{\textstyle{2\over 3}}\, z \,+
{\textstyle{1\over 9}}\, z^2
 + \cdots \cr
g^{(4)}_4(z) &&= \,\,\,\,2 +
 3\, z + \,3  \,
z^2  +\cdots
\een
In this presentation, all computations are manifestly
real. In addition all $g^{(4)\prime}_i(0)$'s are positive as expected and
we simply
take their positive square roots in evaluating \refb{cans} with $f_i$
replaced by $g_i$. 
We get:
\be
\label{cans2}
g^2 S\,\Bigl|_{t^4}= -\, {(-4)(-3) + (-6)(-1)\over 12\cdot
3} \, t^4 = -{1\over 2} \, t^4
\ee
This agrees with the result of the disk computation.

\end{document}